\documentclass[journal]{IEEEtran}
\usepackage[utf8]{inputenc}
\usepackage{graphics,epsfig,epstopdf}
\usepackage{url,amsmath,amssymb}
\usepackage{array}
\usepackage{color,subfigure}
\usepackage{booktabs}
\usepackage{threeparttable}
\usepackage{multirow}
\usepackage{tabularx}
\usepackage{verbatim}
\usepackage[table]{xcolor}
\usepackage{hyperref} 
\usepackage{booktabs}
\usepackage[numbers]{natbib}
\usepackage{amsfonts,amssymb}
\usepackage{graphicx}
\usepackage{float}
\usepackage{enumitem}
\usepackage{subfigure}
\usepackage{filecontents}
\usepackage{algorithm}
\usepackage{algpseudocode}
\usepackage{bm}
\usepackage{xcolor}
\usepackage{xpatch}

\begin{document}

\title{Semantic Importance-Aware Based for Multi-User Communication Over MIMO Fading Channels}

\author{Haotai Liang,
        Zhicheng Bao,
        Wannian An,
        Chen Dong*,
        Xiaodong Xu,~\IEEEmembership{Senior Member,~IEEE,}

\thanks{Haotai Liang, Zhicheng Bao, and Wannian An are with the State Key Laboratory of Networking and Switching Technology, Beijing University of Posts and Telecommunications, Beijing, China (e-mail: lianghaotai@bupt.edu.cn; zhicheng\_bao@bupt.edu.cn; anwannian2021@bupt.edu.cn).}
\thanks{*Chen Dong is the corresponding author and with the State Key Laboratory of Networking and Switching Technology, Beijing University of Posts and Telecommunications, Beijing, China (e-mail: dongchen@bupt.edu.cn).}
\thanks{Xiaodong Xu is with the State Key Laboratory of Networking and Switching Technology, Beijing University of Posts and Telecommunications, Beijing, China, and also with the Department of Broadband Communication, Peng Cheng Laboratory, Shenzhen, Guangdong, China (e-mail: xuxiaodong@bupt.edu.cn).
}
}
\maketitle

%
\begin{abstract}
Semantic communication, as a novel communication paradigm, has attracted the interest of many scholars, with multi-user, multi-input multi-output (MIMO) scenarios being one of the critical contexts. This paper presents a semantic importance-aware based communication system (SIA-SC) over MIMO Rayleigh fading channels. Combining the semantic symbols' inequality and the equivalent subchannels of MIMO channels based on Singular Value Decomposition (SVD) maximizes the end-to-end semantic performance through the new layer mapping method. For multi-user scenarios, a method of semantic interference cancellation is proposed. Furthermore, a new metric, namely semantic information distortion (SID), is established to unify the expressions of semantic performance, which is affected by channel bandwidth ratio (CBR) and signal-to-noise ratio (SNR). With the help of the proposed metric, we derived performance expressions and Semantic Outage Probability (SOP) of SIA-SC for Single-User Single-Input Single-Output (SU-SISO), Single-User MIMO (SU-MIMO), Multi-Users SISO (MU-MIMO) and Multi-Users MIMO (MU-MIMO) scenarios. Numerical experiments show that SIA-SC can significantly improve semantic performance across various scenarios. 
\end{abstract}
\section{Introduction}
Semantic communication systems are considered a potentially promising next-generation communication paradigm, which has gained significant attention in recent years. According to the previous work about the semantic communication system, the new paradigm differs from classical communication systems by jointly considering the representation of semantic information in the source and the impact of the channel on the semantics \cite{56118, ZHANG202260}. 

The end-to-end semantic communication system targeted at specific sources was proposed in the last few years, validating that semantic communication exhibits impressive semantic performance compared to traditional communication, especially under low signal-to-noise (SNR) conditions \cite{8723589, 9398576, 9954279, 9791398}. Due to the joint consideration of channel effects on the semantic performance of the source in the optimization objective of the end-to-end semantic communication systems, the cliff effect issue, which exists in traditional communication systems, has been resolved. In addition, almost all the semantic communication systems designed are not based on the hard decoding schemes of traditional communication systems. Therefore, the mechanism of how the channel affects the performance of semantic communication systems becomes one of the potential directions that require attention. 

Some prior works have noticed and utilized the inequality of semantic symbols to achieve variable length coding schemes for semantic communications. The authors \cite{9791398, bao2023mdvsc} used a hyperprior network \cite{balle2018variational} to fit the probability distribution of semantic signals and performed entropy estimation for each semantic symbol. The magnitude of the entropy expresses the importance of each semantic symbol. Consequently, it becomes possible to select the semantic symbols that need to be transmitted based on their importance level, according to the communication rate. In the research conducted by D. Huang et al., \cite{9953076}, the semantic features were classified into various categories, and the quantization level was adjusted individually for each category. B. Zhang et al. \cite{zhang2023semantic} designed a universal variable-length semantic and channel coding module that can be utilized in different semantic communication systems. By introducing some proxy functions, the rate allocation scheme is learned in an end-to-end manner. 

The paper mentioned above utilized the inequality of semantic symbols and achieved a variable-length rate semantic communication system tailored for single-antenna setups. This paper considers the organic combination of the inequality of semantic symbols with the characteristics of multiple parallel subchannels in Multiple-Input Multiple-Output (MIMO) channels as a necessary and intriguing research direction. MIMO systems have garnered significant interest following the pioneering works of \cite{telatar1999capacity} and \cite{foschini1998limits}. These systems transmit parallel data streams over a MIMO channel, leading to a linear increase in the Shannon capacity with the minimum number of transmit and receive antennas. Compared to single-input single-output (SISO) systems, MIMO systems exhibit a substantial boost in spectral efficiency, known as spatial multiplexing gain. There have recently been semantic communication systems over MIMO fading channels \cite{10026795, wu2022vision, yao2022versatile}. H. Wu et al. \cite{wu2022vision} proposed a Joint Source and Channel Coding (JSCC) scheme based on a Vision Transformer (ViT) for wireless image transmission through MIMO systems, namely ViT-MIMO.  ViT-MIMO can adaptively learn the feature mapping and power allocation on the source image and channel conditions. S. Yao et al. \cite{yao2022versatile} present a novel versatile semantic coded transmission (SCT) scheme over MIMO fading channels named VST-MIMO. An adaptive spatial multiplexing (ASM) module is designed to guide the rate allocation and stream mapping, coupling the source semantics and channel states. However, these works have relied on strong manual assumptions in the rate allocation scheme, while the optimal coding scheme may vary from one transmission task to another. Furthermore, when confronted with changes in the number of antennas, the end-to-end systems over MIMO fading channels will experience variations in their optimal coding schemes, necessitating retraining the entire model, which introduces additional burdens.

Based on the considerations above, we designed a layer mapping scheme based on semantic importance, which integrates the inequality of semantic symbols with the parallel subchannels of the MIMO channel. This scheme can be applied to any end-to-end semantic communication system trained with a single antenna, and it does not require additional modules or impose any extra training burden. 

Furthermore, SIA-SC is combined with Orthogonal Model Division Multiple Access (O-MDMA) \cite{OMDMA, MDMA} technology to extend it to multi-user broadcasting scenarios. From Orthogonal Model Division Multiple Access (O-MDMA), it can be inferred that the semantic signals generated by different semantic models do not result in greater interference within the semantic domain. In addition, a semantic cancellation method is proposed to improve the semantic performance of multiple users.

To provide a more comprehensive explanation of the gains achieved by the SIA-SC, a new metric, namely semantic information distortion (SID), is established to unify the expression of semantic performance \cite{9953095, 9763856, li2023non}, which is affected by channel bandwidth ratio (CBR) \cite{9953110, 9066966} and signal-to-noise ratio (SNR). With the help of the proposed metric, we derived performance expression and Semantic Outage Probability (SOP) of SIA-SC for Single-User Single-Input Single-Output (SU-SISO), Single-User MIMO (SU-MIMO), Multi-Users SISO (MU-SISO) and Multi-Users MIMO (MU-MIMO) scenarios.

The following is a summary of the contributions made by this paper:
\begin{itemize} 
\item A semantic importance-aware based semantic communication system over fading MIMO channel (SIA-SC) is proposed, which integrates the inequality of semantic symbols with the parallel subchannels of the MIMO channel. The SIA-SC is combined with Orthogonal Model Division Multiple Access (O-MDMA) to extend it to a multi-user transmission system in broadcasting scenarios.
\item The performance of a single-antenna semantic communication system based on the entropy model was theoretically analyzed. A unified expression was derived from the perspectives of the CBR and SNR to quantify the system's performance. Based on this unified expression for single-antenna and single-user scenarios (SU-SISO), the performance expression and Semantic Outage Probability (SOP) for the SIA-SC was derived in Single-User MIMO (SU-MIMO), Multi-Users SISO (MU-MIMO) and Multi-Users MIMO (MU-MIMO) scenarios.
\item A large number of numerical and comparative experiments were conducted on SIA-SC. The numerical results demonstrate that our SIA-SC significantly improves the transmission quality compared to the traditional source-channel separation coding scheme using the BPG image compression algorithm with capacity-achieving channel transmission. 
\end{itemize}

The rest of this paper is arranged as follows: 
The framework of the semantic importance-aware based communication system (SIA-SC) is introduced in Section \uppercase\expandafter{\romannumeral2}. Section \uppercase\expandafter{\romannumeral3} analyzes the performance of the SIA-SC, including the definition of the semantic information distortion (SID) and analysis of SIA-SC for SU-SISO, SU-MIMO, MU-SISO, and MU-MIMO scenarios. Section \uppercase\expandafter{\romannumeral4} provides numerical results and corresponding discussions. Finally, conclusions about our work are drawn in section \uppercase\expandafter{\romannumeral5}.

Notation: $\mathbb{R}^{m \times n}$ represent sets of real matrices of size $m\times n$. Variables with uppercase letters in bold-font represent matrices, variables with lowercase letters in bold-font represent vectors. In particular, $\mathbf{S}_i[m,n]$ represent the elements with index $m, n$ in the $i^{th}$ matrix and $\mathbf{s}_i[m]$ represent the elements with index $m$ in the $i^{th}$ vector. $(\cdot)^H$ denote the Hermitian.

\begin{figure*}
    \centering
    \setlength{\abovecaptionskip}{0.cm}
    \includegraphics[width=0.95\textwidth]{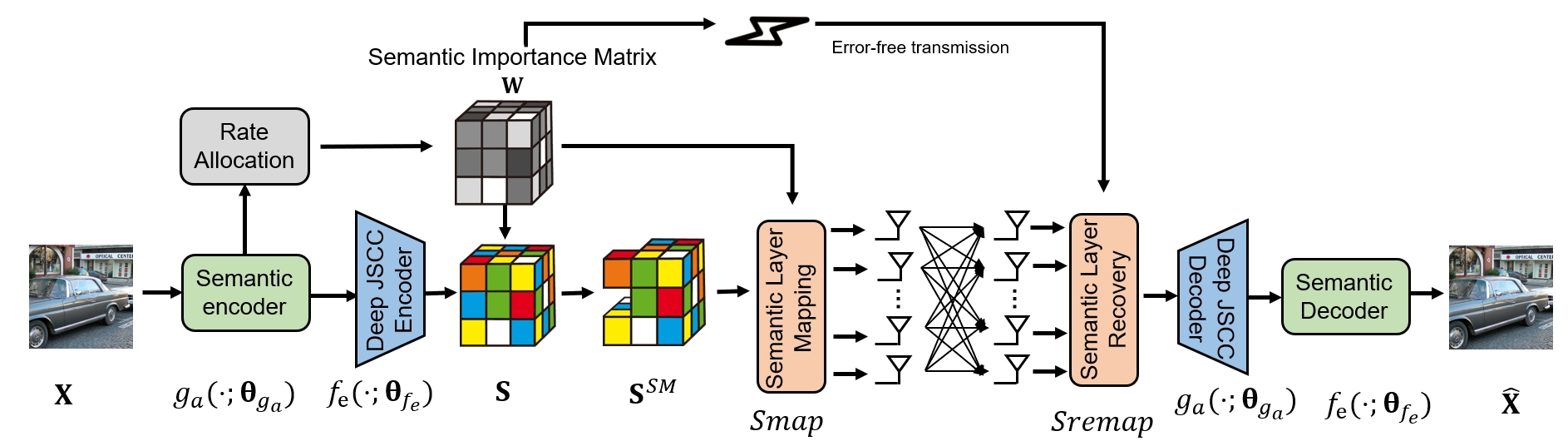}
    \centering
    \caption{The framework of the Semantic Importance-aware semantic communication system (SIA-SC) over fading MIMO channel. The source $\mathbf{X}$ passes through the semantic encoder and DeepJSCC to obtain the semantic symbols $\mathbf{S}$. The important semantic symbol $\mathbf{S}^{SM}$ for transmission is selected by the importance matrix $\mathbf{W}$. Then, through semantic layer mapping, the semantic symbol $\mathbf{S}^{SM}$ is placed on the corresponding antennas. At the receiver, corresponding inverse transformations are performed. The error-free transmission channel is used to ensure the transmission of the importance matrix $\mathbf{W}$, referring to Qin’s design \cite{zhang2023semantic}.}
    \label{SIA-SC}
\end{figure*}
\begin{figure*}
    \centering
    \setlength{\abovecaptionskip}{0.cm}
    \includegraphics[width=0.95\textwidth]{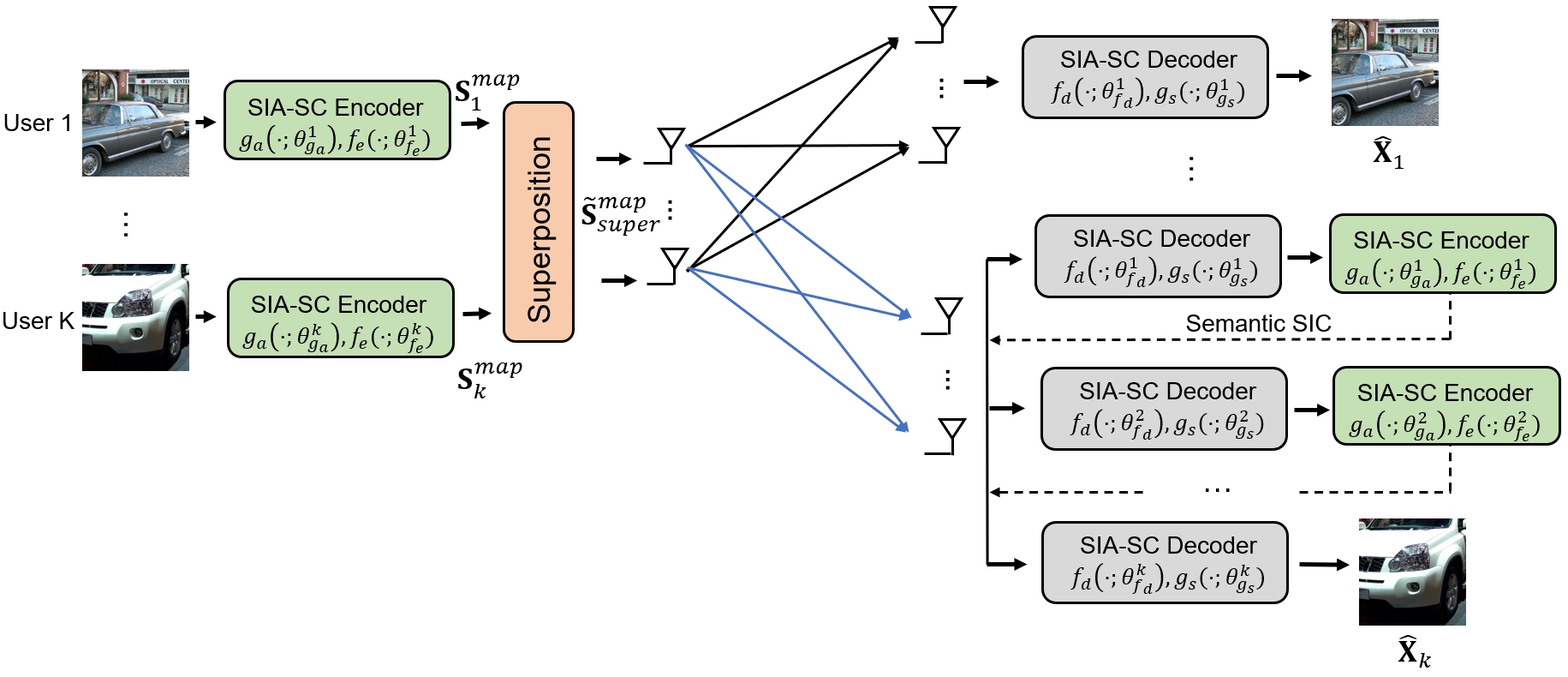}
    \centering
    \caption{The framework of O-MDMA-based Semantic Importance-aware semantic communication system (SIA-SC) over fading MIMO channel for multi-users. Each user independently uses the SIA-SC encoder to obtain semantic transmission symbols $\mathbf{S}_1^{map}$, then superimposes them with unequal power allocation to obtain $\mathbf{\tilde{S}}^{map}_{super}$. At the receiver, each user first decodes the information of the user allocated with higher power and then reduces interference from other users through semantic serial interference cancellation methods.}
    \label{SIA-SC MU}
\end{figure*}

\section{System Model}
This section introduces the semantic importance-aware semantic communication system (SIA-SC) over the fading MIMO channel. Then, the SIA-SC for the multi-user degraded broadcast scenarios is described in detail.

\subsection{The framework of SIA-SC}
A semantic importance-aware communication system is depicted in Fig. \ref{SIA-SC}. Consider a $N_t \times N_r$ MIMO communication system, where a $N_t$ antennas transmitter aims to send an image $\mathbf{X} \in \mathbb{R}^{W\times H\times C}$ to a $N_r$ antennas receiver. The source $\mathbf{X}$ is transformed into latent representation $\mathbf{L}$ and semantic symbols $\mathbf{S}\in \mathbb{R}^{N\times1}$ through semantic encoder and deep JSCC encoder \cite{bao2023mdvsc}, 
\begin{equation}
    \mathbf{L}=g_a(\mathbf{X}; \bm{\theta_{g_a}}),
\end{equation}
\begin{equation}
    \mathbf{S}=f_e(\mathbf{L}; \bm{\theta_{f_e}}),
\end{equation}
where $g_a$ and $f_e$ represent the semantic encoder with the training parameter $\bm{\theta_{g_a}}$ and Deep JSCC with the training parameter $\bm{\theta_{f_e}}$. Before transmitting the semantic symbols, an entropy model is applied to estimate the information entropy of the symbols,  namely semantic importance matrix $\mathbf{W}$, 
\begin{equation}
    \mathbf{W}=p_e(\mathbf{L}; \bm{\theta_{p_e}}),
\end{equation}
where $p_e$ denotes the entropy estimation function \cite{ballé2018variational, 9953110, bao2023mdvsc} with the training parameter $\bm{\theta_{p_e}}$. Details about the entropy estimation function $p_e$ have been detailed in our previous paper \cite{bao2023mdvsc}. The value of the $\mathbf{W}$ can be used to generate a 0-1 mask $\mathbf{M}$ that controls the semantic rate. The semantic symbols $\mathbf{S}^{SM}\in \mathbb{R}^{K\times1}$ to be processed can be represented as, 
\begin{equation}
    \mathbf{S}^{SM}= dropout(\mathbf{S}\odot\mathbf{M}),
\label{dropout}
\end{equation}
where $\odot$ represents the dot product, the zero-value will be discarded to meet certain constraints through the $dropout$ function.
The channel bandwidth ratio (CBR) \cite{9066966} can be denoted as $R\triangleq \frac{K}{W\times H\times C}$.

Different from the semantic communications over fading MIMO channel \cite{wu2022vision, yao2022versatile}, MIMO channels are not introduced into the end-to-end training. Instead, the Singular Value Decomposition (SVD) pre-coding and post-coding schemes are used to decompose the MIMO channels into multiple equivalent sub-channels. Specifically, given the Channel State Information (CSI), the MIMO channel matrix $\mathbf{H}\in \mathbb{R}^{N_t \times N_r}$ is subjected to SVD as follows,
\begin{equation}
    \mathbf{H}= \mathbf{U}\mathbf{\Sigma}\mathbf{V}^H,
\end{equation}
where $\mathbf{U}\in \mathbb{R}^{N_t \times N_t}$ and $\mathbf{V}\in \mathbb{R}^{N_r \times N_r}$ are unitary matrices, and
$\mathbf{\Sigma}\in \mathbb{R}^{N_t \times N_r}$ is a diagonal matrix with its singular values arranged in descending order. In our derivation, we assume that $N_t=N_r=r$, which physically means that the number of antennas at the transmitter is equal to that at the receiver. The $\mathbf{\Sigma}$ is denoted by $\rm{diag}(\sigma_1, \sigma_2, ..., \sigma_r)$, where $\sigma_1\geq \sigma_2\geq ...\geq\sigma_r$. The MIMO channel matrix can be effectively represented as multiple sub-channels with varying channel gains through SVD decomposition. By allocating different semantic symbols based on their importance for semantic performance, it is possible to maximize semantic performance.

\begin{figure*}
    \centering
    \setlength{\abovecaptionskip}{0.cm}
    \includegraphics[width=0.95\textwidth]{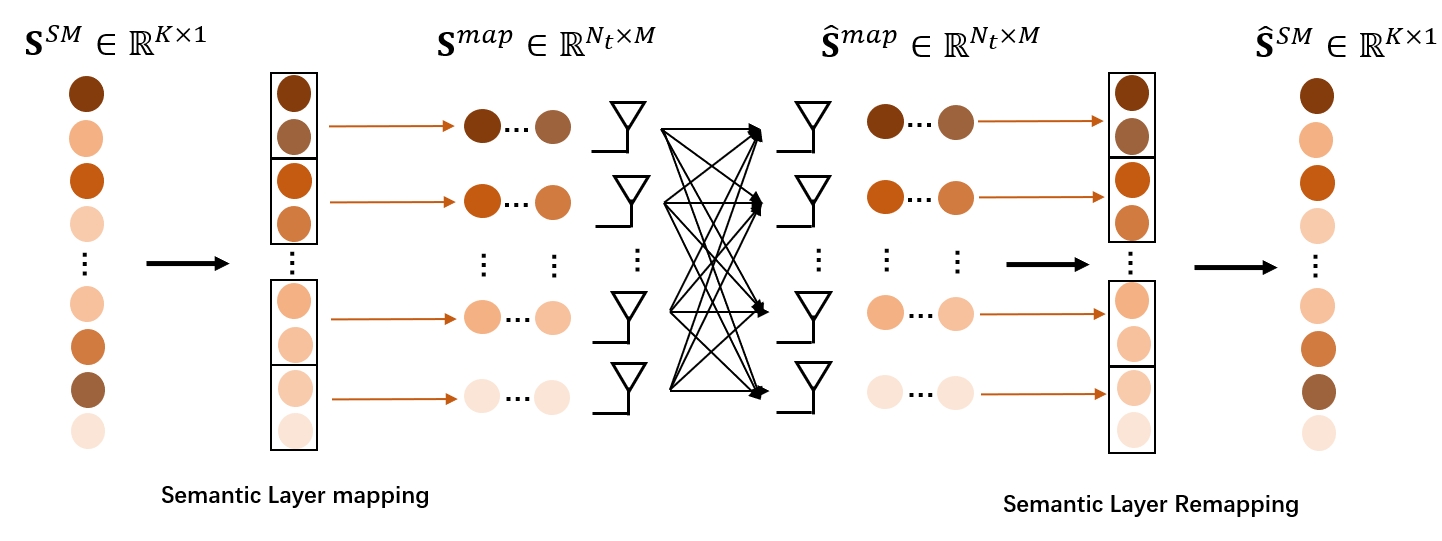}
    \centering
    \caption{Schematic diagram of the specific operation of semantic layer mapping. the semantic transmission symbols $\mathbf{S}^{SM}$ are sorted according to the semantic importance matrix $\mathbf{W}$. Then, the sorted semantic symbols are one-by-one mapped to the antennas from top to bottom.
}
    \label{semantic_mapping}
\end{figure*}

As shown in the left part of Fig. \ref{semantic_mapping}, the importance of each symbol in the semantic symbols $\mathbf{S}^{SM}$ are in random order. The transmitter first sorts $\mathbf{S}^{SM}$ according to the value of the entropy matrix $\mathbf{W}$ and then places it onto the corresponding antennas, which is called semantic layer mapping. The mapped transmission symbols $\mathbf{S}^{map}$ through semantic layer mapping can be represented as,
\begin{equation}
\mathbf{S}^{map} = Smap(\mathbf{S}^{SM};\mathbf{W}), 
\end{equation}
where $\mathbf{S}^{map}\in \mathbb{R}^{N_t \times M}$ and $M=\frac{K}{N_t}$ denote the number of symbols in the semantic layer for each antenna. 

With SVD precoding, $\mathbf{S}^{map}$ will be transformed to $\mathbf{\tilde{S}}^{map}$,
\begin{equation}
    \mathbf{\tilde{S}}^{map}= \mathbf{V}\mathbf{S}^{map}.
\end{equation}
$\mathbf{\tilde{S}}^{map}$ is transmitted to the receiver through the fading MIMO channel and can be represented as, 
\begin{equation}
    \mathbf{Y}= \mathbf{H}\mathbf{\tilde{S}}^{map}+\mathbf{n},
\label{receiver_signal}
\end{equation}
where $\mathbf{H}$ follows independent and identically distributed (i.i.d.) complex Gaussian distribution with zero mean and variance $\sigma_h^2$, i.e., $\mathbf{H}[i, j]\sim \mathcal{CN}(0, \sigma_h^2)$, $\mathbf{n}$ follows i.i.d. complex Gaussian distribution with zero mean and variance $\sigma_n^2$, i.e., $\mathbf{n}[i, j]\sim \mathcal{CN}(0, \sigma_n^2)$.
In the receiver, premultiply by the matrix
$\mathbf{U}^H$, 
\begin{equation}
\label{postprocess_signal}
    \mathbf{\tilde{Y}}= \mathbf{\Sigma}\mathbf{S}^{map}+\mathbf{\tilde{n}},
\end{equation}
where $\mathbf{\tilde{n}}$ still follows i.i.d. complex Gaussian distribution. 

As shown in the right part of Fig. \ref{semantic_mapping}, the receiver uses the $\mathbf{W}$ to remap the received semantic symbols $\mathbf{\tilde{Y}}$ back to the semantic symbols of the original sequence $\mathbf{\hat{S}}^{SM}$,
\begin{equation}
\mathbf{\hat{S}}^{SM} = Sremap(\mathbf{\tilde{Y}};\mathbf{W}). 
\end{equation}
Finally, the image $\mathbf{\hat{X}}$ and the latent representation $\mathbf{\hat{L}}$ are restored through the Deep JSCC decoder $f_d$ and semantic decoder $g_s$,
\begin{equation}
\mathbf{\hat{L}}=f_d(\mathbf{\mathbf{\hat{S}}}^{SM}; \bm{\theta}_{f_d}),
\end{equation}
\begin{equation}
\mathbf{\hat{X}}=g_s(\mathbf{\hat{L}}; \bm{\theta}_{g_s}).
\end{equation}

\subsection{The framework of O-MDMA-based SIA-SC for multi-users}
As shown in Fig. \ref{SIA-SC MU}, K users use their respective SIA-SC encoders $g_a(\cdot;\theta_{g_a}^k), f_e(\cdot;\theta_{f_e}^k)$ to extract semantic transmission symbols $S_k$ and perform layer mapping according to their respective semantic importance matrix $\mathbf{W}_k$ to obtain sorted semantic symbols $\mathbf{S}^{map}_k$. 

From O-MDMA\cite{OMDMA, MDMA}, it can be inferred that the semantic signals generated by different semantic models do not result in greater interference within the semantic domain. Under the interference cancellation capabilities of DeepJSCC, it is possible to recover the specified semantic performance effectively. Similarly, in the digital domain, the semantic symbols of different users are combined with unequal power,
\begin{equation}
\mathbf{\tilde{S}}^{map}_{super}=\sum_{k=1}^{K}\sqrt{P_k}\mathbf{V}_k\mathbf{S}^{map}_k,
\end{equation}
where $P_k (P_1>...>P_k>...>P_K)$ is the power allocated by the $k^{th}$ user, and $\mathbf{V}_k$ is the right singular matrix of the transmission channel for the $k^{th}$ user.
The semantic signal received by user $k$ can be represented as follows,
\begin{equation}
    \mathbf{Y}_k= \mathbf{H}_k\mathbf{\tilde{S}}^{map}_{super}+\mathbf{n}_k.
\end{equation}
First, decode the user's received semantic signal with the highest allocated power as same as Non-Orthogonal Multiple Access \cite{7842433, 7117391},
\begin{equation}
\begin{aligned}
    \mathbf{\tilde{Y}}_1&= \sqrt{P_1}\mathbf{U}_1^H(\mathbf{H}_1\mathbf{\tilde{S}}^{map}_{super}+\mathbf{n}_1)
    \\&=\sqrt{P_1}\mathbf{\Sigma}_1\mathbf{S}^{map}_1+\sum_{k=2}^K\sqrt{P_k}\mathbf{\Sigma}_1\mathbf{V}_1^H\mathbf{V}_k\mathbf{S}^{map}_k + \mathbf{\tilde{n}}.
\end{aligned}
\label{MU-MIMO-user1}
\end{equation}
Then, the error caused by the interference of the last two terms on semantic performance is eliminated as much as possible by the ability of DeepJSCC,
\begin{equation}
\mathbf{\hat{X}}_1=g_s(f_d(Sremap(\mathbf{\tilde{Y}}_1;\mathbf{W}_1);\bm{\theta}_{f_d}^1);\bm{\theta}_{g_s}^1).
\end{equation}
User 2 performs the same operations as User 1 to obtain $\mathbf{\hat{X}}_1$ and eliminate User 1's interference,
\begin{equation}
\mathbf{\hat{S}}^{map}_1=Smap(f_e(g_a(\mathbf{\hat{X}}_1;\bm{\theta}_{g_s}^1);\bm{\theta}_{f_e}^1);\mathbf{W}_1).
\end{equation}
Assuming the pilot signals are shared among the users, User 2 can obtain  the signal with User 1's interference canceled $\mathbf{\tilde{S}}^{map}_{super,2}$,
\begin{equation}
\mathbf{\tilde{S}}^{map}_{super,2}=\sum_{k=2}^{K}\sqrt{P_k}\mathbf{V}_k\mathbf{S}^{map}_k + \triangle_1,
\label{error}
\end{equation}
where $\triangle_1=|\sqrt{P_1}\mathbf{V}_1(\mathbf{S}^{map}_1-\mathbf{\hat{S}}^{map}_1)|$. Similar to User 1, the post-processed received signal of User 2 can be represented as follows,
\begin{equation}
\begin{aligned}
    \mathbf{\tilde{Y}}_2&= \mathbf{U}_2^H(\mathbf{H}_2\mathbf{\tilde{S}}^{map}_{super, 2}+\mathbf{n}_2)
    \\&=\sqrt{P_2}\mathbf{\Sigma}_2\mathbf{S}^{map}_2+\sum_{k=3}^K\sqrt{P_k}\mathbf{\Sigma}_2\mathbf{V}_2^H\mathbf{V}_k\mathbf{S}^{map}_k + \tilde{\triangle}_1 + \mathbf{\tilde{n}},
\end{aligned}
\label{MU-MIMO-user2}
\end{equation}
where $\tilde{\triangle}_1 = \mathbf{\Sigma}_2\mathbf{V}_2^H\triangle_1$. In general, for User $i$, 
\begin{equation}
\mathbf{\tilde{Y}}_i=\mathbf{\Sigma}_i\mathbf{S}^{map}_i+\sum_{k=i+1}^K\sqrt{P_k}\mathbf{\Sigma}_i\mathbf{V}_i^H\mathbf{V}_k\mathbf{S}^{map}_k + \tilde{\triangle}_{i-1} + \mathbf{\tilde{n}}.
\end{equation}

\begin{figure*}[h]
\centering
\begin{minipage}{0.3\linewidth}
\centerline{\includegraphics[width=\textwidth]{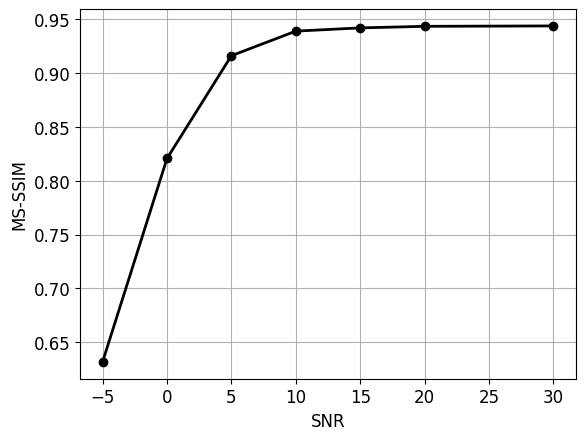}}
\centerline{(a) SNR-SID-MSSSIM}
\end{minipage}
\begin{minipage}{0.3\linewidth}
\centerline{\includegraphics[width=\textwidth]{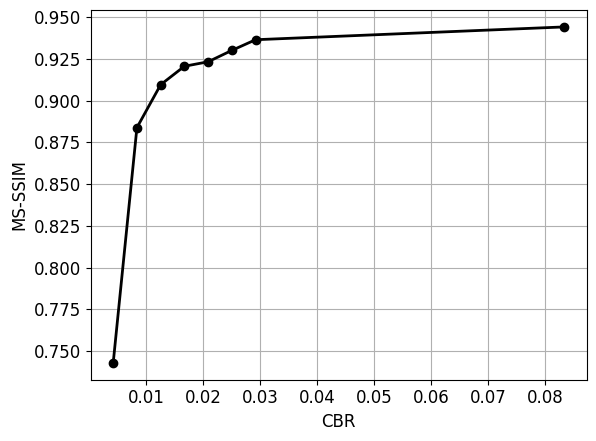}}
\centerline{(b) CBR-SID-MSSSIM}
\end{minipage}
\begin{minipage}{0.3\linewidth}
\centerline{\includegraphics[width=\textwidth]{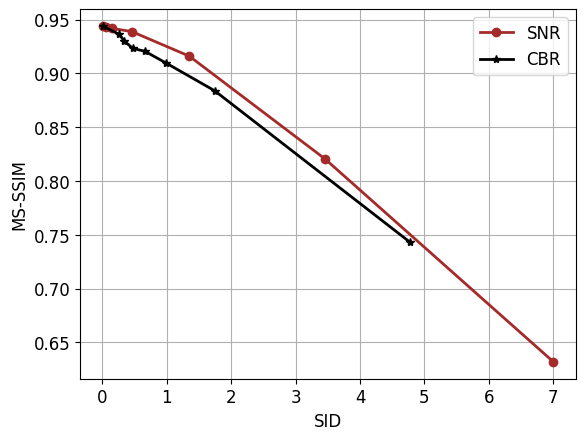}}
\centerline{(c) SID-MSSSIM}
\end{minipage}
\caption{The influence of SNR and CBR on
the SID, and their impact on semantic performance. As SNR and CBR decrease, SID increases, and semantic performance decreases. The impact of SID generated by SNR and CBR on semantic performance (MS-SSIM) is consistent.}
\label{SISO_SDR}
\end{figure*}


\begin{figure*}
    \centering
    \setlength{\abovecaptionskip}{0.cm}
    \includegraphics[width=0.95\textwidth]{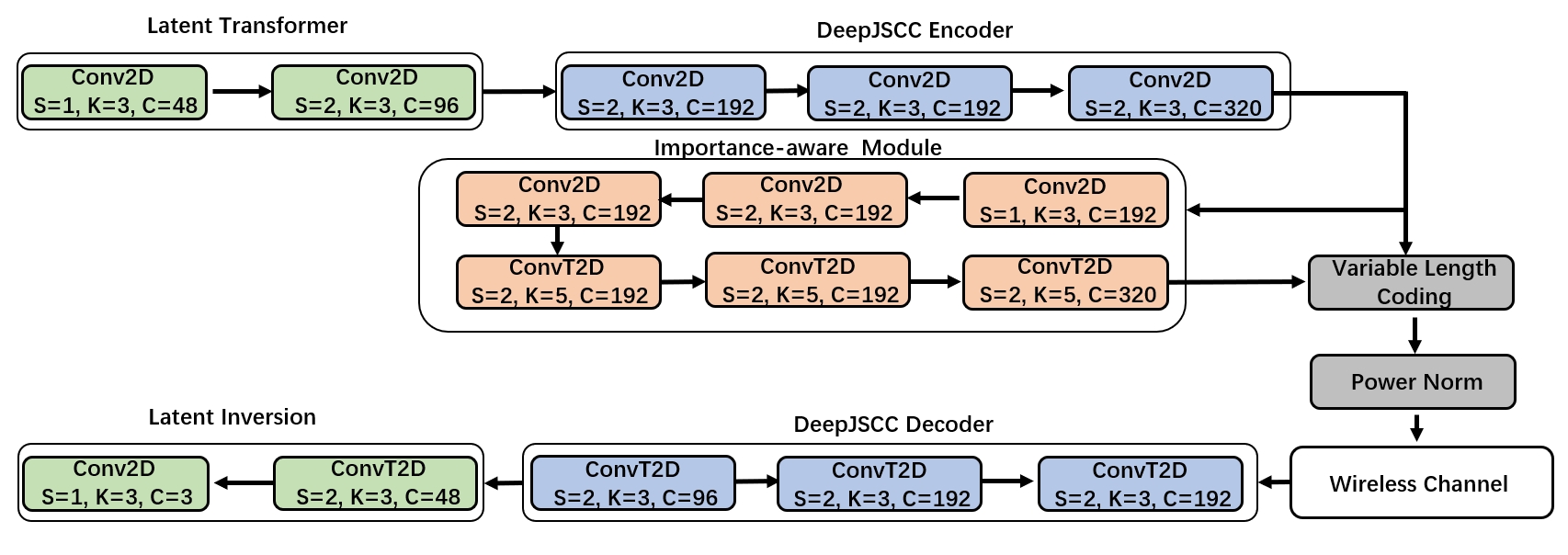}
    \caption{The specific network architecture for SIA-SC. The network architecture is mainly composed of a Latent Transformer, DeepJSCC Encoder, DeepJSCC Decoder, Latent Inversion and the importance-aware module, where parameter $S$ represents stride, $K$ represents kernel size, and $C$ represents the number of channels.}
    \label{model_architecture}
\end{figure*}

\section{Semantic System Analysis}
The key focus of the SIA-SC system introduced in the previous section is the semantic symbol inequality, meaning that each semantic symbol contributes differently to semantic performance. Currently, existing performance analyses for semantic communication \cite{9953095, 9763856, li2023non} have not considered the inequality of semantic symbols. 

This section first introduces semantic information distortion (SID) and Semantic Outage Probability (SOP). A unified expression is provided to represent the impact of SNR and CBR on semantic performance. Subsequently, we further analyze the performance in MU-SISO, SU-MIMO, and MU-MIMO scenarios.

\subsection{Semantic Information Distortion and Semantic Outage Probability}
X. Mu et al. \cite{9953095} first propose to employ the data regression method and approximate the semantic performance concerning the received SNR. However, semantic systems with different semantic encoding lengths (or expressed using CBR) require using functions with different parameters to express their performance curves.

It is easy to realize that a key difference between semantic communication and traditional communication lies in the fact that the receiver no longer performs hard/soft decisions; instead, it directly feeds the contaminated transmission symbols into the semantic decoder for semantic recovery. So, the indicative metric influenced by the SNR is no longer the symbol error rate but rather the degree of distortion in the transmitted symbols. Assuming that the remaining semantic symbols are transmitted through the AWGN channel to the receiver, 
\begin{equation}
    \hat{\mathbf{s}}[i] = \mathbf{s}[i] + n,\ 0<i<K+1,
\end{equation}
\begin{equation}
SID(\gamma)=\sum_{i=1}^{K}\mathbf{w}[i]\mathbb{E}[|\hat{\mathbf{s}}[i]-\mathbf{s}[i]|^2]=\sum_{i=1}^{K}\mathbf{w}[i]\sigma_n^2,
\label{SISO_gama}
\end{equation}
\begin{equation}
\sum_{i=1}^N\mathbf{w}[i]=1, \quad
\mathbb{E}[\mathbf{s}^*\mathbf{s}]\leq P,
\end{equation}
where $\gamma$ represents the SNR of the received signal and  $\mathbf{w}[i] (\mathbf{w}[1]>...>\mathbf{w}[i]>...>\mathbf{w}[K]>...>\mathbf{w}[N])$ represents the degree of importance of the $i^{th}$ semantic symbol $\mathbf{s}[i]$.  $\sum_{i=1}^N\mathbf{w}[i]|\mathbf{s}[i]-\hat{\mathbf{s}}[i]|$ is named as SID.

Furthermore, current semantic communications with adaptive rate control achieve their adaptation by setting relatively less important semantic symbols to zero, such as Nonlinear Transform Source-Channel Coding for Semantic Communications (NTSCC) \cite{9791398}, Wireless Model Division Video Semantic Communication (MDVSC) \cite{bao2023mdvsc} and the SIA-SC proposed in the second section of this paper. 
Considering the end-to-end semantic communication systems, controlling CBR is typically achieved by zeroing out semantic symbols as Eq. (\ref{dropout}), which can be equivalently expressed as,
\begin{equation}
SID(CBR)=\sum_{i=K+1}^{N}\mathbf{w}[i]|\mathbf{s}[i]-0|^2=\sum_{i=K+1}^{N}\mathbf{w}[i]|\mathbf{s}[i]|^2.
\label{SISO_cbr}
\end{equation}
We have separately simulated the influence of SNR and CBR on the SID, and their impact on semantic performance is depicted in Fig. \ref{SISO_SDR} using MDVSC. As shown in Fig. \ref{SISO_SDR}(a)(b), as SNR and CBR decrease, SID increases and the semantic performance decreases. More importantly, as shown in Fig. \ref{SISO_SDR}(c), the impact of SID generated by SNR and CBR on semantic performance (MS-SSIM) is consistent.

Therefore, the performance expressions \cite{9953095} can be rewritten  as follows:
\begin{equation}
\xi(CBR, \gamma)=\xi(SID)=\xi(\sum_{i=1}^N\mathbf{w}[i]|\mathbf{s}[i]-\mathbf{\hat{s}}_i|^2).
\label{SID_su_siso}
\end{equation}
The estimation expressions for the SID in the SU-SISO scenarios can be obtained from Eq. (\ref{SISO_gama}) and Eq. (\ref{SISO_cbr}) as follows:
\begin{equation}
SID_{SU-SISO}(\gamma, CBR)=\sum_{i=1}^K\mathbf{w}[i]\sigma_n^2 + \sum_{i=K+1}^N\mathbf{w}[i]|\mathbf{S}[i]|^2.
\label{siso_su_sid}
\end{equation}

With the help of the defined and derived SID, the SOP can be defined as the probability of exceeding the distortion constraint $SID_{th}$, which is expressed as \cite{zhang2023scan},
\begin{equation}
P\{SID(\gamma, CBR)>SID_{th}\}.
\label{SOP_define}
\end{equation}
When considering a SU-SISO system, according to Eq. (\ref{siso_su_sid}) and Eq. (\ref{SOP_define}), where only Gaussian noise $\mathbf{n}$ is a random variable, then the semantic outage probability is equivalent to,
\begin{equation}
\begin{aligned}
P_{SU-SISO}&=P\{\sum_{i=1}^K\mathbf{w}[i]\mathbf{n}^2[i]\\&>SID_{th}-\sum_{i=K+1}^N\mathbf{w}[i]|\mathbf{s}[i]|^2\},
\end{aligned}
\label{SOP_su_siso}
\end{equation}
where $\sum_{i=1}^K\mathbf{w}[i]\mathbf{n}[i]$ is the sum of squares of Gaussian distributions with different variances, following a generalized non-central chi-squared distribution for which there is currently no closed-form expression. Fortunately, due to the largeness of $K$, i.e., the degrees of freedom of the non-central chi-squared distribution are large, it further approximates a Gaussian distribution according to the Central Limit Theorem,
\begin{equation}
\sum_{i=1}^K\mathbf{w}[i]\mathbf{n}[i]\sim\mathcal{C}(\sigma_n^2\sum_{i=1}^K\mathbf{w}[i], 2\sigma_n^4\sum_{i=1}^K\mathbf{w}^2[i]).
\label{Gaussian_sop}
\end{equation}
By utilizing the cumulative distribution function of the Gaussian distribution, the SOP can be calculated.

\subsection{SU-MIMO scenarios}
The SIA-SC MIMO system mentioned in Section \uppercase\expandafter{\romannumeral2} uses SVD precoding and postcoding methods. According to Eq. (\ref{receiver_signal}), the gains of each sub-channel are $(\sigma_1, \sigma_2, ..., \sigma_r)$ respectively. The SNR of the $j^{th}$ equivalent sub-channel can be expressed as,
\begin{equation}
\gamma_j=\frac{\sigma_jP}{\sigma_n}.
\end{equation}
The expected value of SID generated by channel noise can be represented as follows:
\begin{equation}
SID(\gamma)=\sum_{j=1}^r\sum_{i=1}^M\mathbf{W}[i,j]\frac{\sigma_n^2}{\sigma_j^2P},
\end{equation}
where $M$ denotes the number of semantic symbols for each sub-channel and $M\times r=K$. The impact of CBR on semantic performance recovery is the same as in the SU-SISO scenario, so the estimation expression for the SU-MIMO scenario can be expressed as,
\begin{equation}
\begin{aligned}
SID_{SU-MIMO}(\gamma, CBR)&=\sum_{j=1}^r\sum_{i=1}^M\mathbf{W}[i,j]\frac{\sigma_n^2}{\sigma_j^2P}\\&+\sum_{i=K+1}^N\mathbf{w}[i]|\mathbf{s}[i]|^2.
\end{aligned}
\label{SID_su_mimo}
\end{equation}
Assuming a time-invariant channel for transmitting an image, relying on Channel State Information at the Transmitter (CSIT), the SOP for SU-MIMO scenarios can be expressed similarly to Eq. (\ref{SOP_su_siso}), as follows:
\begin{equation}
\begin{aligned}
P_{SU-MIMO}&=P\{\sum_{j=1}^r\sum_{i=1}^M\mathbf{W}[i,j]\frac{\mathbf{n}^2[i,j]}{\sigma_j^2P}\\&>SID_{th}-\sum_{i=K+1}^N\mathbf{w}[i]|\mathbf{s}[i]|^2\}.
\end{aligned}
\label{SOP_su_mimo}
\end{equation}
Similarly to Eq. (\ref{Gaussian_sop}), $\sum_{j=1}^r\sum_{i=1}^M\mathbf{W}[i,j]$ approximately follows a Gaussian distribution,
\begin{equation}
\begin{aligned}
\sum_{j=1}^r\sum_{i=1}^M\mathbf{W}[i,j]&\frac{\mathbf{n}^2[i,j]}{\sigma_j^2P}\sim
\\&\mathcal{C}(\sigma_n^2\sum_{j=1}^r\sum_{i=1}^M\frac{\mathbf{W}[i,j]}{\sigma_j^2P}, 2\sigma_n^4\sum_{j=1}^r\sum_{i=1}^M(\frac{\mathbf{W}[i,j]}{\sigma_j^2P})^2).
\label{Gaussian_sop_su_mimo}
\end{aligned}
\end{equation}

\subsection{MU-SISO scenarios}
Considering a scenario with two users in a downlink, where the base station allocates power $P_1$ to user 1 and power $P_2$ to user 2, with $P_1 > P_2$, 
\begin{equation}
\mathbf{s}_{super}=\sqrt{P_1}\mathbf{s}_1+\sqrt{P_2}\mathbf{s}_2.
\end{equation}
The semantic signals received by user 1 and user 2 respectively are
\begin{equation}
\begin{aligned}
\mathbf{y}_1&=\mathbf{h}_1\mathbf{s}_{super}+\mathbf{n}_1\\&=\mathbf{h}_1\sqrt{P_1}\mathbf{s}_1+\mathbf{h}_1\sqrt{P_2}\mathbf{s}_2+\mathbf{n}_1,
\end{aligned}
\end{equation}
\begin{equation}
\begin{aligned}
\mathbf{y}_2&=\mathbf{h}_2\mathbf{s}_{super}+\mathbf{n}_2\\&=\mathbf{h}_2\sqrt{P_1}\mathbf{s}_1+\mathbf{h}_2\sqrt{P_2}\mathbf{s}_2+\mathbf{n}_2,
\end{aligned}
\end{equation}
where $\mathbf{n}_i\sim \mathcal{CN}(0, \sigma_{n, i}^2)$. First, decode the semantic signal of user 1, 
\begin{equation}
\gamma_1=\frac{P_1|\mathbf{h}_1|^2}{\sigma_{n,1}^2+P_2|\mathbf{h}_1|^2},
\end{equation}
\begin{equation}
\mathbf{\hat{s}}_1=\mathbf{s}_1+\sqrt{\frac{\rho_2}{\rho_1}}\mathbf{s}_2+\frac{\mathbf{n}_1}{\mathbf{h}_1\sqrt{P_1}}.
\end{equation}
The SID expectation affected by the channel $\gamma$ and CBR can be expressed as,
\begin{equation}
\begin{aligned}
SID_{MU-SISO, 1}&(\gamma, CBR)=\\&\sum_{i=1}^K\mathbf{w}_1[i]\left[\frac{P_2}{P_1}+\frac{\sigma^2}{|h_1|^2P_1}\right]\\&+\sum_{i=K+1}^N\mathbf{w}_1[i]|\mathbf{s}_1[i]|^2.
\label{SID_mu_siso_1}
\end{aligned}
\end{equation}

Next, according to Fig. \ref{SIA-SC MU}, the decoder for user 1 is utilized to decode the image of user 1, 
\begin{equation}
\begin{aligned}
\mathbf{\hat{x}}_{1, sic}&=\\&g_s(f_d(Sremap(\mathbf{s}_1+\sqrt{\frac{P_2}{P_1}}\mathbf{s}_2+\frac{\mathbf{n}_1}{\mathbf{h}_2\sqrt{P_1}};\\&\mathbf{W}_1);\bm{\theta}_{f_d}^1);\bm{\theta}_{g_s}^1),
\end{aligned}
\end{equation}

\begin{equation}
\begin{aligned}
\mathbf{\hat{s}}_{1, sic}=Smap(f_e(g_a(\mathbf{\hat{x}}_{1,sic};\mathbf{W}_1);\bm{\theta}_{g_s}^1);\bm{\theta}_{f_e}^1);\mathbf{W}_1).
\end{aligned}
\end{equation}
As the receiver does not use hard/soft decisions, it has to rely on its own understanding ability to derive a message as close as possible to the original message. But this reconstruction of $\mathbf{s}_1$ cannot be perfect. Furthermore, $\sum_{i=1}^K\mathbf{w}_2[i]|\mathbf{s}_1[i]-\mathbf{s}_{1,sic}[i]|^2$ is related to the performance of decoding the user 1 image first, i.e. related to $SID_{MU-SISO, 1}$.  Thus, $SID_{sic}=\sum_{i=1}^K\mathbf{w}_2[i]|\mathbf{s}_1[i]-\mathbf{s}_{1,sic}[i]|^2$ can be expressed by $SID_{MU-SISO, 1}$,
\begin{equation}
SID_{sic}=\sum_{i=1}^K\mathbf{w}_2[i]|\mathbf{s}_1[i]-\mathbf{s}_{1,sic}[i]|^2 = te(SID_{MU-SISO, 1}),
\label{SID_TE}
\end{equation}
where $te$ denote the model can be established for $SID_{sic}$ and $SID_{MU-SISO, 1}$. 

User 2 then utilizes an estimate of $\mathbf{\hat{s}}_{1,sic}$ to eliminate interference from User 1,
\begin{equation}
\mathbf{\hat{s}}_{2}=\frac{\mathbf{y}_2-\mathbf{h}_2\sqrt{P_1}\hat{\mathbf{s}}_{1,sic}}{\mathbf{h}_2\sqrt{P_2}}.
\end{equation}
\begin{equation}
\begin{aligned}
SID_{MU-SISO, 2}(\gamma, CBR)= &te(SID_1)+\sum_{i=1}^K\mathbf{w}_2[i]\frac{\sigma^2}{|h_2|^2P_2}\\&+\sum_{i=K+1}^N\mathbf{w}_2[i]|\mathbf{s}_2[i]|^2.
\end{aligned}
\label{SID_mu_siso_2}
\end{equation}

The SOP for User 1 and User 2 $P_{MU-SISO, 1}$, $P_{MU-SISO, 2}$ can be expressed as,
\begin{equation}
\begin{aligned}
&P_{MU-SISO, 1}=P\{\sum_{i=1}^K\mathbf{w}_1[i]\frac{\mathbf{n}^2[i]}{|h_1|^2P_1}\\&>SID_{th}-\sum_{i=1}^K\mathbf{w}_1[i]\frac{P_2}{P_1}-\sum_{i=K+1}^N\mathbf{w}_1[i]|\mathbf{s}_1[i]|^2\},
\end{aligned}
\label{SOP_mu_siso_1}
\end{equation}

\begin{equation}
\begin{aligned}
\sum_{i=1}^K\mathbf{w}_1[i]&\frac{\mathbf{n}^2[i]}{|h_1|^2P_1}\sim
\\&\mathcal{C}(\sigma_n^2\sum_{i=1}^K\frac{\mathbf{w}_1[i]}{|h_1|^2P_1}, 2\sigma_n^4\sum_{i=1}^K(\frac{\mathbf{w}_1[i]}{|h_1|^2P_1})^2).
\label{Gaussian_sop_mu_siso}
\end{aligned}
\end{equation}

\begin{equation}
\begin{aligned}
&P_{MU-SISO, 2}=P\{\sum_{i=1}^K\mathbf{w}_2[i]\frac{\mathbf{n}^2[i]}{|h_2|^2P_2}\\&>SID_{th}-te(SID_{MU-SISO, 1})-\sum_{i=K+1}^N\mathbf{w}_2[i]|\mathbf{s}_2[i]|^2\},
\end{aligned}
\label{SOP_mu_siso_2}
\end{equation}

\begin{equation}
\begin{aligned}
\sum_{i=1}^K\mathbf{w}_2[i]&\frac{\mathbf{n}^2[i]}{|h_2|^2P_2}\sim
\\&\mathcal{C}(\sigma_n^2\sum_{i=1}^K\frac{\mathbf{w}_2[i]}{|h_2|^2P_2}, 2\sigma_n^4\sum_{i=1}^K(\frac{\mathbf{w}_2[i]}{|h_2|^2P_2})^2).
\label{Gaussian_sop_mu_siso_2}
\end{aligned}
\end{equation}

\subsection{MU-MIMO scenarios}
Considering a scenario with two users in a downlink, where the base station allocates power $P_{1}$ to user 1 and power $P_{2}$ to user 2, with $P_{1} > P_{2}$. Similar to Eq. (\ref{MU-MIMO-user1}),
\begin{equation}
\begin{aligned}
    \mathbf{\tilde{Y}}_1&= \sqrt{P_{1}}\mathbf{\Sigma}_1\mathbf{S}^{map}_1+\sqrt{P_{2}}\mathbf{\Sigma}_1\mathbf{V}_1\mathbf{V}_2\mathbf{S}^{map}_2 + \mathbf{\tilde{n}},
\end{aligned}
\end{equation}
The SID expectation affected by the channel and CBR for User 1 can be expressed as,
\begin{equation}
\begin{aligned}
    &SID_{MU-MIMO,1}=\sum_{j=1}^r\sum_{i=1}^M\mathbf{W}_1[i,j]\Bigg[\frac{\sigma_n^2}{\sigma_j^2P_1}+\frac{P_2}{P_1}\Bigg]\\&+\sum_{i=K+1}^N\mathbf{w}_1[i]|\mathbf{s}^{map}_1[i]|^2.
\end{aligned}
\label{mu_mimo_sid_1}
\end{equation}

The following discusses the SID of the imperfect semantic SIC decoding for user 2. As same as Eq. (\ref{MU-MIMO-user2}),
\begin{equation}
\begin{aligned}
\mathbf{\tilde{Y}}_2&=\sqrt{P_2}\mathbf{\Sigma}_2\mathbf{S}^{map}_2+\sqrt{P_1}\mathbf{\Sigma}_2\mathbf{V}_2^H\mathbf{V}_1|\mathbf{S}_1^{map}-\hat{\mathbf{S}}_1^{map}| + \mathbf{\tilde{n}}.
\end{aligned}
\end{equation}
Consistent with the discussion in the previous subsection, $\tilde{\triangle}_1 = \mathbf{\Sigma}_2\mathbf{V}_2^H\mathbf{V}_1|\mathbf{S}_1^{map}-\hat{\mathbf{S}}_1^{map}|$ is related to $SID_{MU-MIMO,1}$, 
\begin{equation}
\sum_{j=1}^r\sum_{i=1}^M\mathbf{W}_2[i,j]\tilde{\triangle}_1^2=te(SID_{MU-MIMO,1}),
\end{equation}
and the SID for User 2 can be given by,
\begin{equation}
\begin{aligned}
    &SID_{MU-MIMO,2}=\sum_{j=1}^r\sum_{i=1}^M\mathbf{W}_2[i,j]\frac{\sigma_n^2}{\sigma_j^2P_2}+\\&\frac{P_1}{P_2}te(SID_{MU-MIMO,1})+\sum_{i=K+1}^N\mathbf{w}_2[i]|\mathbf{s}^{map}_2[i]|^2.
\end{aligned}
\label{mu_mimo_sid_2}
\end{equation}
The SOP for User 1 and User 2 $P_{MU-MIMO, 1}$, $P_{MU-MIMO, 2}$ can be expressed as,
\begin{equation}
\begin{aligned}
&P_{MU-MIMO, 1}=P\{\sum_{j=1}^r\sum_{i=1}^M\mathbf{W}_1[i,j]\frac{\mathbf{n}^2[i,j]}{\sigma_j^2P_1}\\&>SID_{th}-\sum_{j=1}^r\sum_{i=1}^M\mathbf{W}_1[i,j]\frac{P_2}{P_1}-\sum_{i=K+1}^N\mathbf{w}_1[i]|\mathbf{s}^{map}_1[i]|^2\},
\end{aligned}
\label{SOP_mu_mimo_1}
\end{equation}

\begin{equation}
\begin{aligned}
\sum_{j=1}^r\sum_{i=1}^M&\mathbf{W}_1[i,j]\frac{\mathbf{n}^2[i,j]}{\sigma_j^2P_1}\sim
\\&\mathcal{C}(\sigma_n^2\sum_{j=1}^r\sum_{i=1}^M\frac{\mathbf{W}_1[i,j]}{\sigma_j^2P_1}, 2\sigma_n^4\sum_{j=1}^r\sum_{i=1}^M(\frac{\mathbf{W}_1[i]}{\sigma_j^2P_1})^2).
\label{Gaussian_sop_mu_mimo}
\end{aligned}
\end{equation}

\begin{equation}
\begin{aligned}
&P_{MU-MIMO, 2}=P\{\sum_{i=1}^K\mathbf{w}_2[i]\frac{\mathbf{n}^2[i]}{|h_2|^2P_2}\\&>SID_{th}-te(SID_{MU-SISO, 1})-\sum_{i=K+1}^N\mathbf{w}_2[i]|\mathbf{s}_2[i]|^2\},
\end{aligned}
\label{SOP_mu_mimo_2}
\end{equation}

\begin{equation}
\begin{aligned}
\sum_{j=1}^r\sum_{i=1}^M&\mathbf{W}_2[i,j]\frac{\mathbf{n}^2}{\sigma_j^2P_2}\sim
\\&\mathcal{C}(\sigma_n^2\sum_{j=1}^r\sum_{i=1}^M\frac{\mathbf{W}_2[i,j]}{\sigma_j^2P_2}, 2\sigma_n^4\sum_{j=1}^r\sum_{i=1}^M(\frac{\mathbf{W}_2[i,j]}{\sigma_j^2P_2})^2).
\label{Gaussian_sop_mu_mimo_2}
\end{aligned}
\end{equation}

\begin{figure*}[h]
\centering
\begin{minipage}{0.32\linewidth}
\centerline{\includegraphics[width=\textwidth]{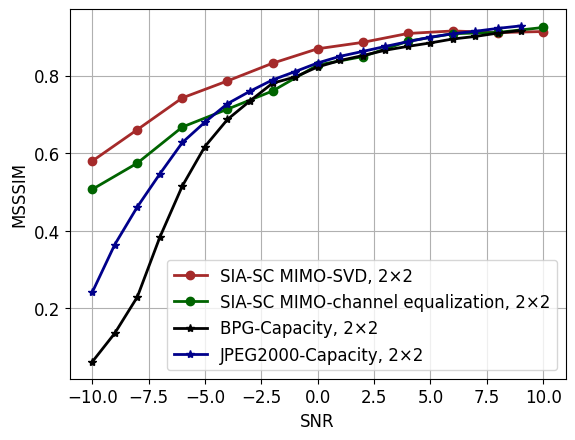}}
\centerline{(a) SNR-MSSSIM, $2\times2$ MIMO}
\end{minipage}
\begin{minipage}{0.32\linewidth}
\centerline{\includegraphics[width=\textwidth]{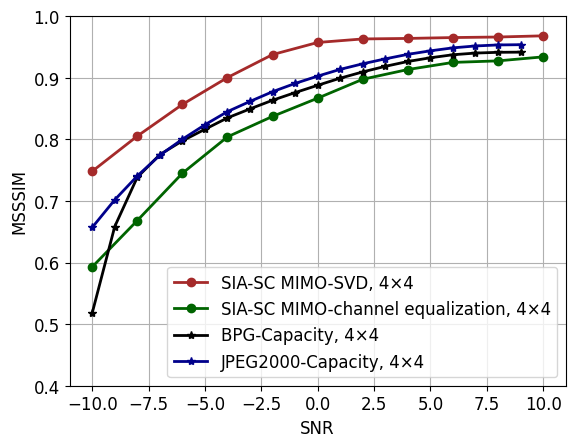}}
\centerline{(b) SNR-MSSSIM, $4\times4$ MIMO}
\end{minipage}
\begin{minipage}{0.32\linewidth}
\centerline{\includegraphics[width=\textwidth]{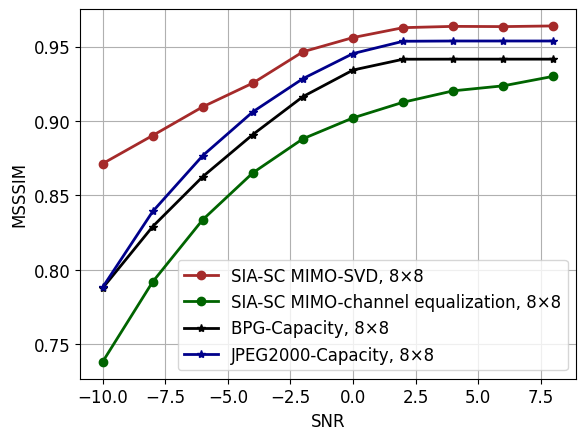}}
\centerline{(c) SNR-MSSSIM, $8\times8$ MIMO}
\end{minipage}
\begin{minipage}{0.32\linewidth}
\centerline{\includegraphics[width=\textwidth]{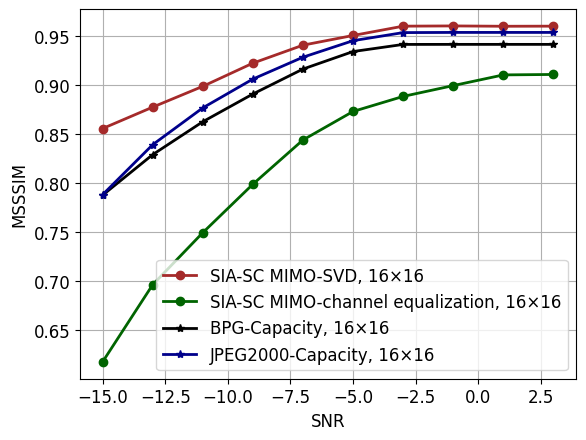}}
\centerline{(d) SNR-MSSSIM, $16\times16$ MIMO}
\end{minipage}
\begin{minipage}{0.32\linewidth}
\centerline{\includegraphics[width=\textwidth]{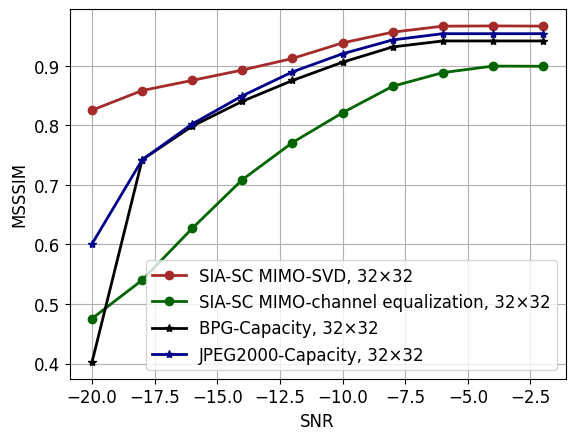}}
\centerline{(e) SNR-MSSSIM, $32\times32$ MIMO}
\end{minipage}
\begin{minipage}{0.32\linewidth}
\centerline{\includegraphics[width=\textwidth]{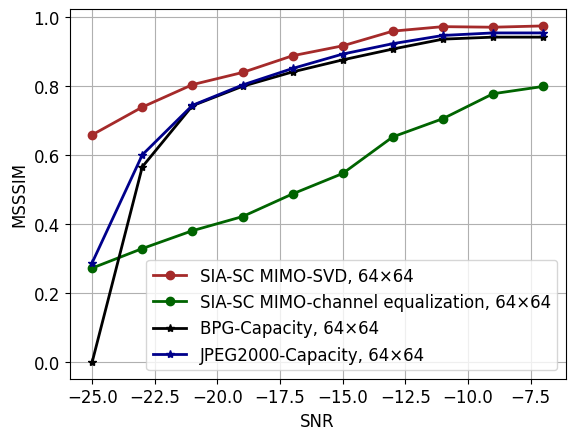}}
\centerline{(f) SNR-MSSSIM, $64\times64$ MIMO}
\end{minipage}
\caption{Single-user Transmission Performance. The SIA-SC MIMO-SVD scheme outperforms the separation-base benchmark in all SNRs. In a large-scale 64x64 MIMO scenario, the SIA-SC MIMO-SVD scheme surpasses the SIA-SC MIMO-channel equalization scheme by over 10 dB.}
\label{SNR-MSSSIM MIMO}
\end{figure*}

\begin{figure}
    \centering
    \setlength{\abovecaptionskip}{0.cm}
    \includegraphics[width=0.45\textwidth]{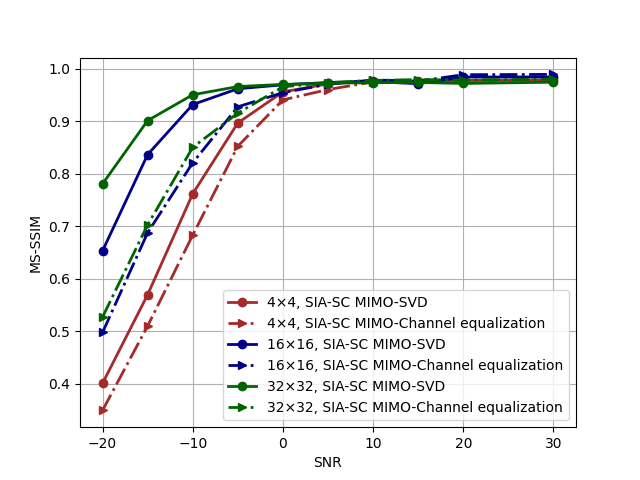}
    \centering
    \caption{Transmission Performance for SIA-SC MIMO-SVD with different numbers of antennas. The advantages of the SIA-SC MIMO-SVD scheme become more significant as the number of antennas increases. }
    \label{eq_svd}
\end{figure}

\begin{figure*}
    \centering
    \setlength{\abovecaptionskip}{0.cm}
    \includegraphics[width=0.95\textwidth]{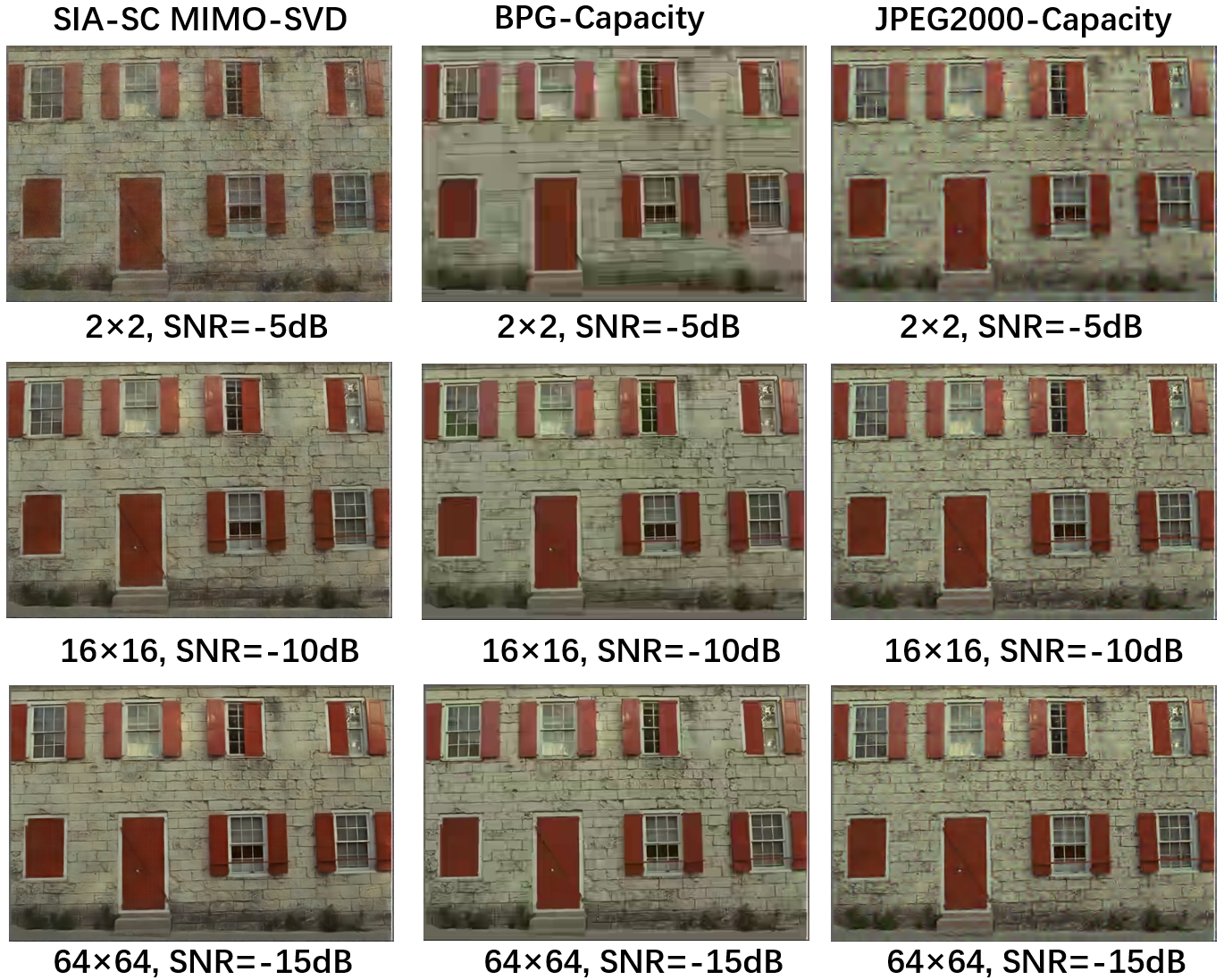}
    \centering
    \caption{Examples of reconstructed images produced by the SIA-SC MIMO-SVD and the baseline schemes (BPG-Capacity and JPEG2000-Capacity). From top to bottom are scenes with different antenna numbers and SNR, and from left to right are the proposed SIA-SC MIMO-SVD scheme, BPG-Capacity scheme, and JPEG2000-Capacity scheme. The visualized examples of the proposed scheme show that the impact of the channel on the image is a global faintening effect, still retaining details, while the comparative schemes show that the impact of the channel on the image is more of a blurring of details.} 
    \label{vis}
\end{figure*}

\begin{figure*}[h]
\centering
\begin{minipage}{0.32\linewidth}
\centerline{\includegraphics[width=\textwidth]{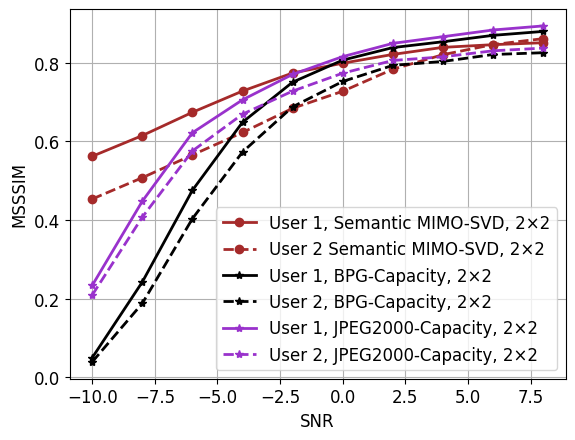}}
\centerline{(a) SNR-MSSSIM, $2\times2$ MU-MIMO}
\end{minipage}
\begin{minipage}{0.32\linewidth}
\centerline{\includegraphics[width=\textwidth]{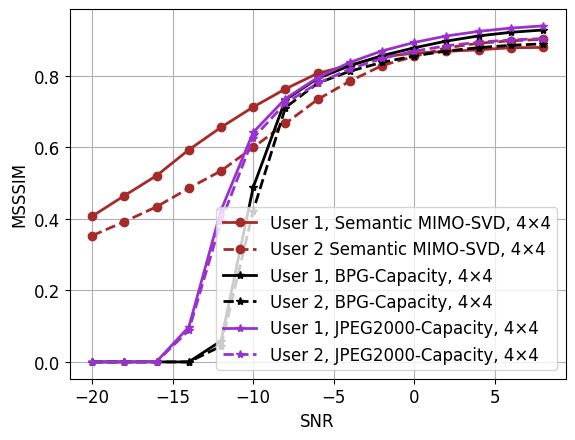}}
\centerline{(b) SNR-MSSSIM, $4\times4$ MU-MIMO}
\end{minipage}
\begin{minipage}{0.32\linewidth}
\centerline{\includegraphics[width=\textwidth]{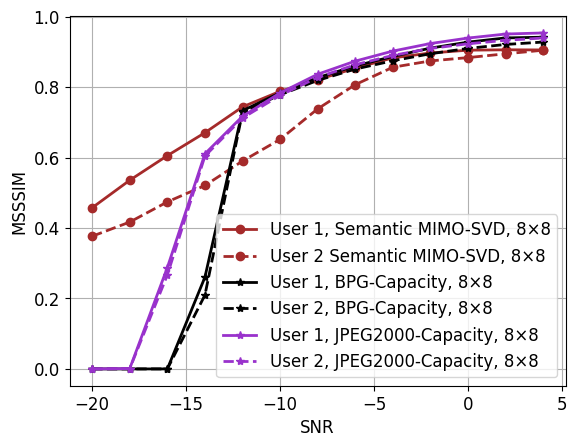}}
\centerline{(c) SNR-MSSSIM, $8\times8$ MU-MIMO}
\end{minipage}
\begin{minipage}{0.32\linewidth}
\centerline{\includegraphics[width=\textwidth]{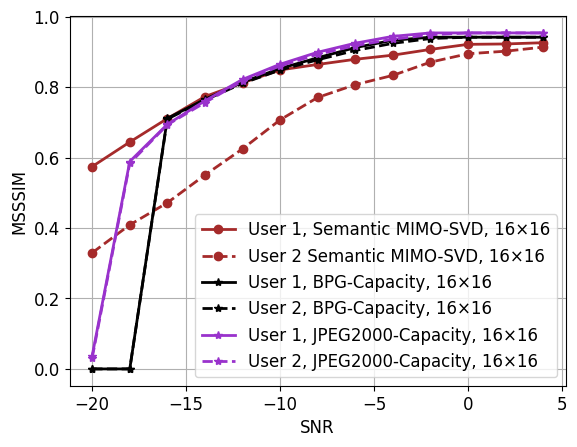}}
\centerline{(d) SNR-MSSSIM, $16\times16$ MU-MIMO}
\end{minipage}
\begin{minipage}{0.32\linewidth}
\centerline{\includegraphics[width=\textwidth]{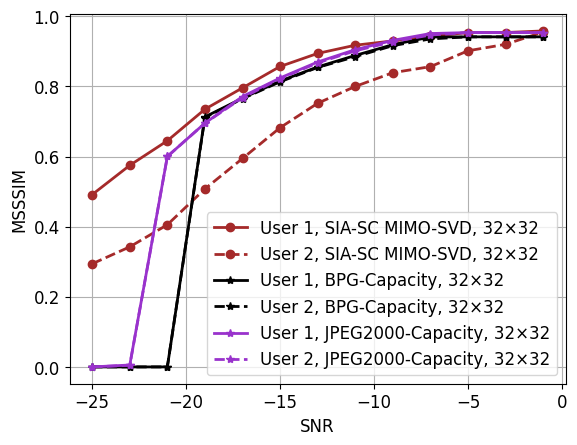}}
\centerline{(e) SNR-MSSSIM, $32\times32$ MU-MIMO}
\end{minipage}
\begin{minipage}{0.32\linewidth}
\centerline{\includegraphics[width=\textwidth]{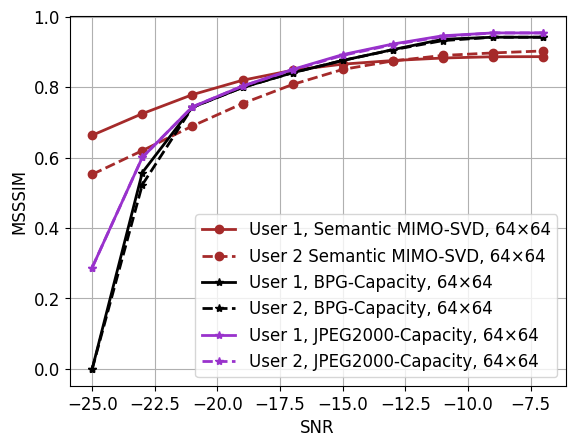}}
\centerline{(f) SNR-MSSSIM, $64\times64$ MU-MIMO}
\end{minipage}
\caption{Muti-user Transmission Performance. The performance of users allocated with higher power in the multi-user multiantenna system is mostly similar to the BPG-Capacity and JPEG2000-Capacity schemes at high SNR. At low SNR, the proposed scheme has a distinct advantage.}
\label{MU SNR-MSSSIM MIMO}
\end{figure*}

\begin{figure}
    \centering
    \setlength{\abovecaptionskip}{0.cm}
    \includegraphics[width=0.45\textwidth]{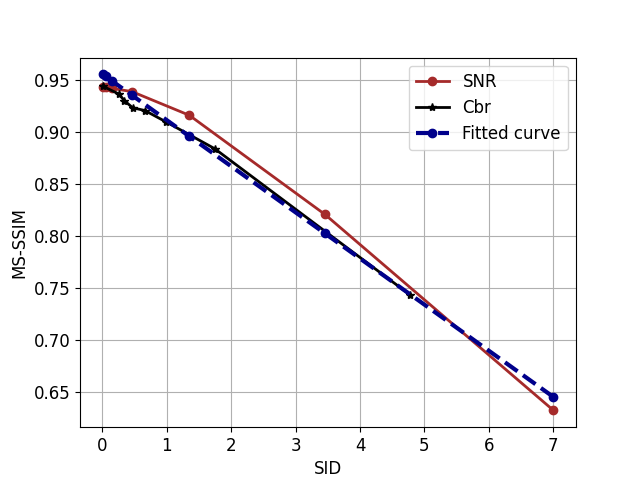}
    \centering
    \caption{The average SID and MS-SSIM values for all test sets are calculated and the linear function is used to fit them.}
    \label{SID_fit}
\end{figure}

\begin{figure}
    \centering
    \setlength{\abovecaptionskip}{0.cm}
    \includegraphics[width=0.45\textwidth]{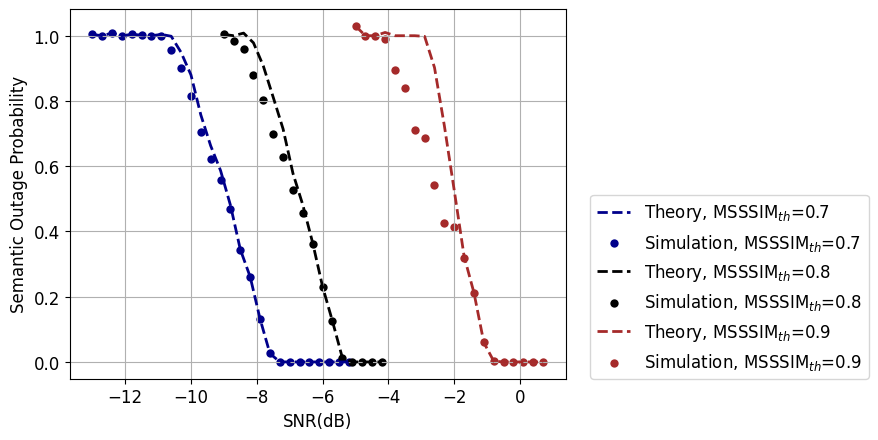}
    \centering
    \caption{The dashed line represents the SOP for the SU-SISO scenario obtained through Eq. (\ref{SOP_su_siso}) and Eq. (\ref{Gaussian_sop}), while the discrete points are derived from simulations, which indicates that Eq. (\ref{SOP_su_siso}) and Eq. (\ref{Gaussian_sop}) provide an approximate closed-form expression for the SOP in the SU-SISO scenario.}
    \label{SOP_SISO_SU}
\end{figure}

\begin{figure*}[h]
\centering
\begin{minipage}{0.32\linewidth}
\centerline{\includegraphics[width=\textwidth]{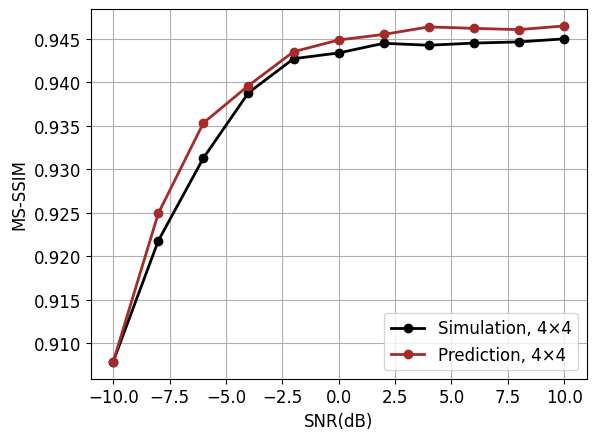}}
\centerline{(a) SNR-MSSSIM, $4\times4$ SU-MIMO}
\end{minipage}
\begin{minipage}{0.32\linewidth}
\centerline{\includegraphics[width=\textwidth]{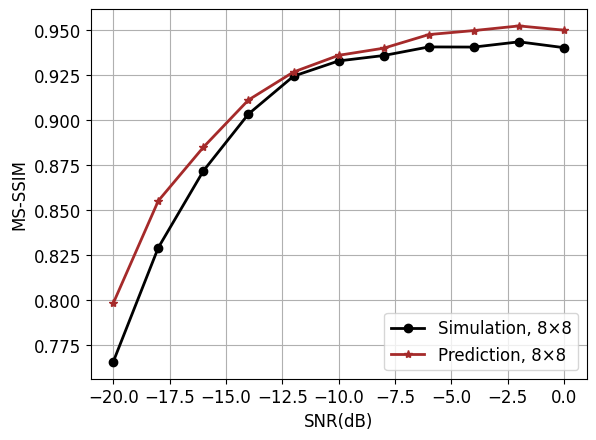}}
\centerline{(b) SNR-MSSSIM, $8\times8$ SU-MIMO}
\end{minipage}
\begin{minipage}{0.32\linewidth}
\centerline{\includegraphics[width=\textwidth]{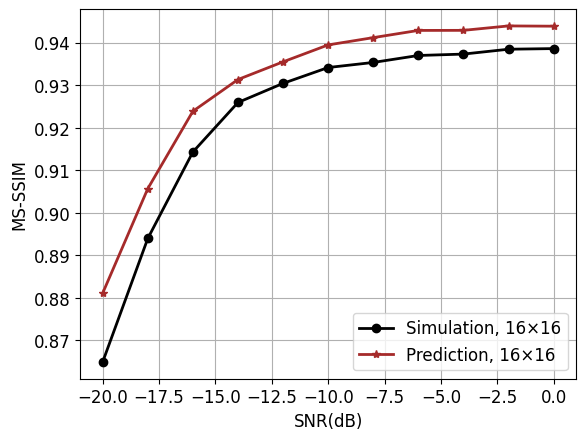}}
\centerline{(c) SNR-MSSSIM, $16\times16$ SU-MIMO}
\end{minipage}
\caption{Validation of Semantic Performance Analysis for SU-MIMO scenario. The predicted values are slightly higher than the simulated values, and the error gradually increases with more antennas. Overall, the transmitter can predict the performance of the image at the receiver given the known SNR.}
\label{SU MIMO predict simulation}
\end{figure*}

\begin{figure}
    \centering
    \setlength{\abovecaptionskip}{0.cm}
    \includegraphics[width=0.45\textwidth]{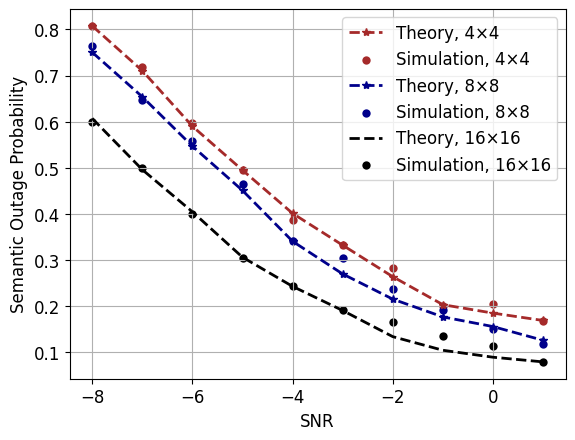}
    \centering
    \caption{The dashed line represents the SOP for the SU-MIMO scenario obtained through Eq. (\ref{SOP_su_mimo}) and Eq. (\ref{Gaussian_sop_su_mimo}). In this simulation, an interruption occurs when the semantic performance of MS-SSIM is less than 0.9.}
    \label{SOP_MIMO_SU}
\end{figure}

\begin{figure*}[h]
\centering
\begin{minipage}{0.32\linewidth}
\centerline{\includegraphics[width=\textwidth]{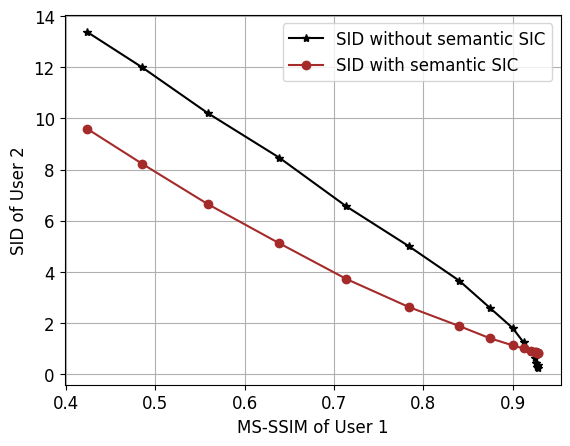}}
\centerline{(a) MS-SSIM of User 1 \textemdash SID of User 2 }
\end{minipage}
\begin{minipage}{0.32\linewidth}
\centerline{\includegraphics[width=\textwidth]{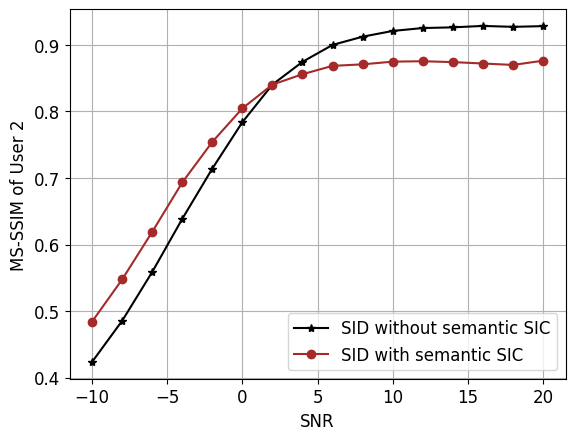}}
\centerline{(b) SNR \textemdash MS-SSIM of User 2}
\end{minipage}
\begin{minipage}{0.32\linewidth}
\centerline{\includegraphics[width=\textwidth]{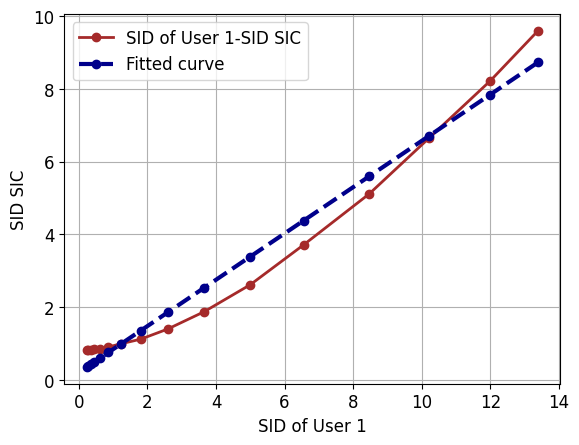}}
\centerline{(c) SID of User 1 \textemdash SID SIC}
\end{minipage}
\caption{(a) The horizontal axis represents User 1’s performance, and the vertical axis represents User 2’s SID. It can be seen that using the semantic SIC scheme can significantly reduce User 2’s SID under poor channel conditions. (b) Through the denoising ability of the semantic model itself, the semantic SIC method can eliminate certain noise. (c) The horizontal axis represents the SIC of User 1 and the vertical axis represents the SID after
semantic SIC.}
\label{without_with_semantic_SIC}
\end{figure*}

\begin{figure}
    \centering
    \setlength{\abovecaptionskip}{0.cm}
    \includegraphics[width=0.45\textwidth]{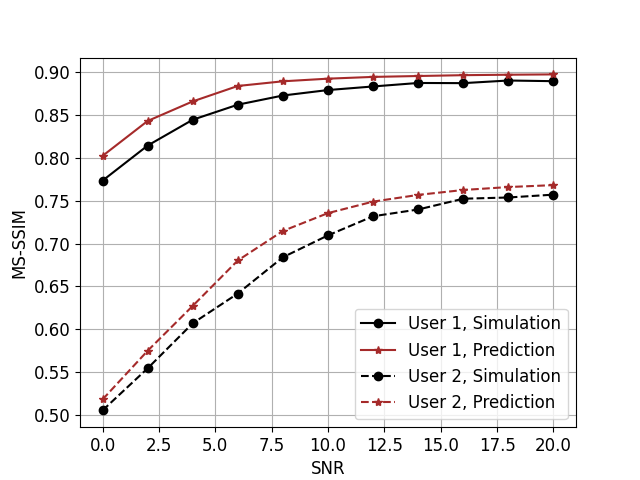}
    \centering
    \caption{Validation of Semantic Performance Analysis for MU-SISO scenario. The predicted semantic performance values are slightly higher than the actual simulated values, but the overall trend is consistent and the gap is not significant.}
    \label{mu_siso}
\end{figure}

\begin{figure}
    \centering
    \setlength{\abovecaptionskip}{0.cm}
    \includegraphics[width=0.40\textwidth]{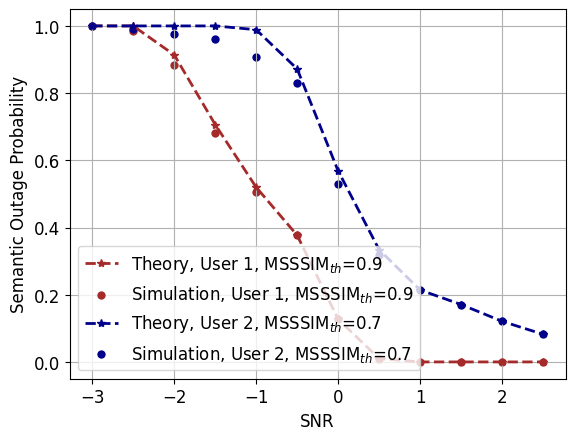}
    \centering
    \caption{The theoretical values and simulation values of the SOP of user 1 and user 2 under the expected performance (MS-SSIM) of 0.9 and 0.7 respectively are plotted. The simulation values are basically on the theoretical curve.}
    \label{SOP_SISO_MU}
\end{figure}

\begin{figure*}[h]
\centering
\begin{minipage}{0.32\linewidth}
\centerline{\includegraphics[width=\textwidth]{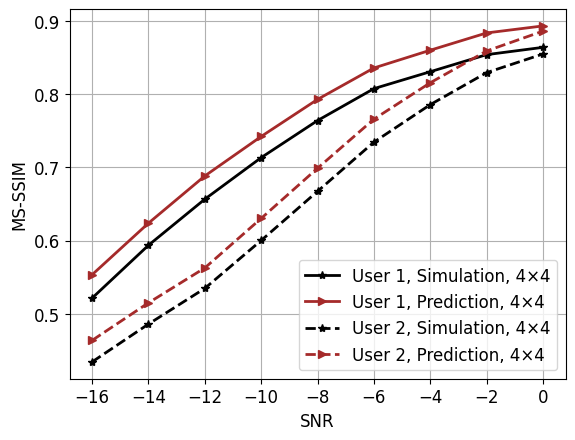}}
\centerline{(a) SNR-MSSSIM, 4$\times$4}
\end{minipage}
\begin{minipage}{0.32\linewidth}
\centerline{\includegraphics[width=\textwidth]{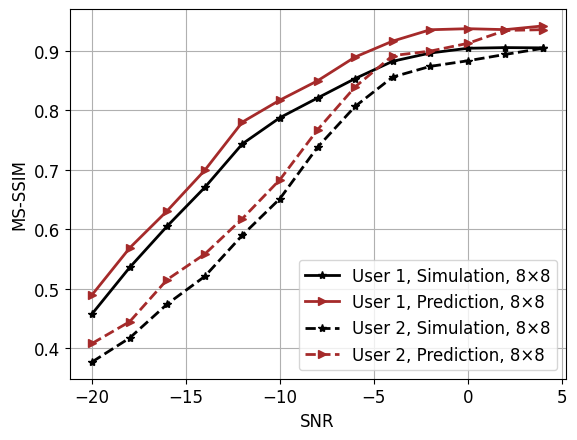}}
\centerline{(b) SNR-MSSSIM, 8$\times$8}
\end{minipage}
\begin{minipage}{0.32\linewidth}
\centerline{\includegraphics[width=\textwidth]{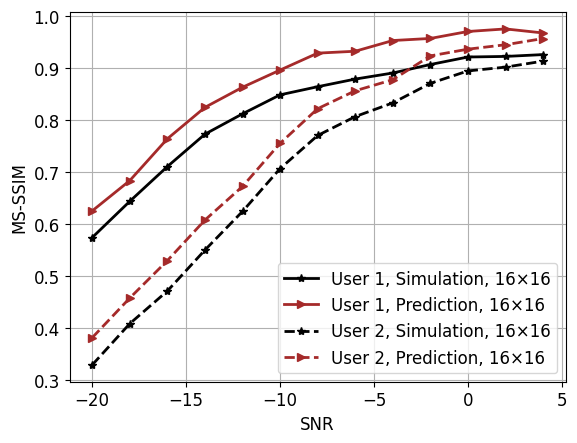}}
\centerline{(c) SNR-MSSSIM, 16$\times$16}
\end{minipage}
\caption{Validation of Semantic Performance Analysis for MU-MIMO scenario. As with other scenarios, the predicted values are slightly greater than the simulated values, with the overall trend remaining consistent and the errors being relatively small.}
\label{MU_MIMO_Simulation_predict}
\end{figure*}

\begin{figure*}[ht]
\centering
\begin{minipage}{0.32\linewidth}
\centerline{\includegraphics[width=\textwidth]{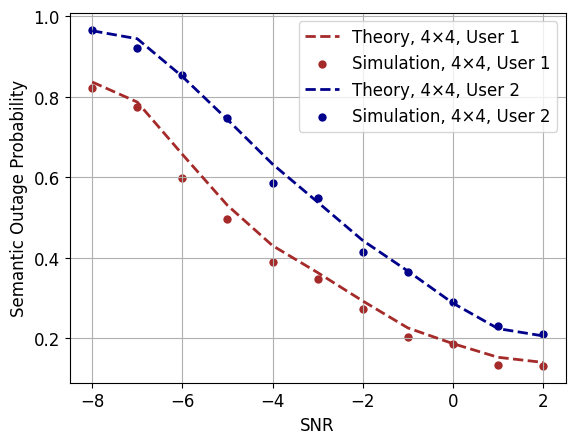}}
\centerline{(a) SNR-SOP, 4$\times$4}
\end{minipage}
\begin{minipage}{0.32\linewidth}
\centerline{\includegraphics[width=\textwidth]{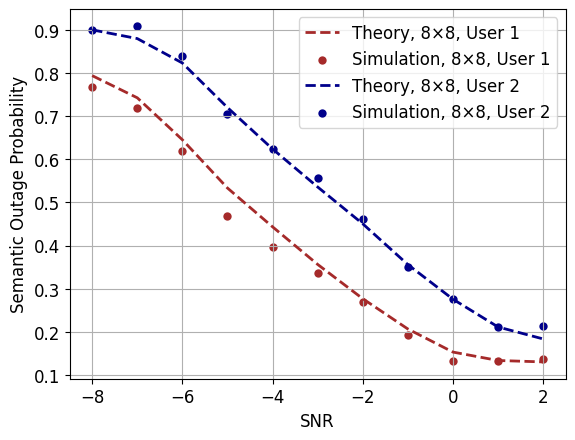}}
\centerline{(b) SNR-SOP, 8$\times$8}
\end{minipage}
\begin{minipage}{0.32\linewidth}
\centerline{\includegraphics[width=\textwidth]{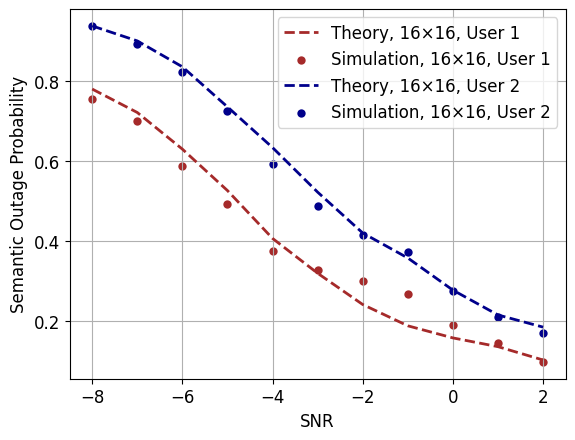}}
\centerline{(c) SNR-SOP, 16$\times$16}
\end{minipage}
\caption{The SOP for the 4x4, 8x8, and 16x16 MU-MIMO scenarios, with an MS-SSIM target value set at 0.9. The simulation values closely align with the theoretical curve, with a slightly larger error observed in the 16x16 scenario, although it remains within an acceptable range.}
\label{SOP_MIMO_MU}
\end{figure*}
\section{Experiments and Discussions}

\subsection{Experimental Setup}
\subsubsection{Datasets}
Our semantic importance-aware based communication system over MIMO Rayleigh fading channels for images is trained on the Open Images datasets \cite{Benenson_2019_CVPR} which has a training set of over 100,000 images, a testing set of over 1000 images. In this paper, the images on the Open Images dataset are downscaled to $512\times512$ for the model training. Furthermore, the Open Images test set and the Kodak dataset ($768\times512$) \cite{kodak} are used for model testing.
\subsubsection{Network architectures}
It's worth emphasizing that our multi-antenna approach applies to other semantic communication systems based on semantic importance, such as Qin's \cite{zhang2023semantic} and Dai's \cite{9791398}. In this section, we have modified our previous MDVSC \cite{bao2023mdvsc} system developed for image-oriented semantic communication based on semantic importance. The training phase is similar to MDVSC, which is treated as an end-to-end semantic communication system for a single antenna. The specific network architecture is illustrated in Fig. \ref{model_architecture}, including a Latent Transformer, DeepJSCC Encoder, DeepJSCC Decoder, Latent Inversion, and Importance-aware module. 
\subsubsection{Baseline}
As a benchmark for comparison, we consider a scheme using JPEG, JPEG2000, and BPG for image compression, together with capacity-achieving channel coding \cite{wu2022vision} (using waterfilling, NOMA capacity) to transmit the compressed images over the channel. 
The source-capacity schemes calculate the MIMO channel capacity, which is the maximum amount of lossless transmission data, and then compress the image to the maximum amount of lossless transmission data to ensure that traditional receivers can successfully decode. 
In addition, the following equation can be obtained by left multiplying the inverse matrix of the diagonal matrix $\mathbf{\Sigma}$ to the Eq. (\ref{postprocess_signal}),
\begin{equation}
\mathbf{\tilde{Y}}_{eq}=\mathbf{\Sigma}^{-1}\mathbf{\tilde{Y}}= \mathbf{S}^{map}+\mathbf{\tilde{n}}.
\end{equation}
This indicates that semantic importance is not distinguished on the MIMO layer mapping, named the SIA-SC channel-equalization scheme.

MS-SSIM \cite{1284395} is employed  as the metric for evaluation, which is defined as follows,
\begin{equation}
     MS-SSIM(x, y) = \frac{(2\mu_{x}\mu_{y}+C_1)(2\sigma_{xy}+C_2)}{(\mu_{x}^2+\mu_{y}^2+C_1)(\sigma_{x}^2+\sigma_{y}^2+C_2)}, 
\end{equation}
where $\mu_x, \sigma_x$ and $\sigma_{xy}$ are the mean, standard deviation, and cross-correlation between the two patches $x$, $y$, respectively. $C_1$ and $C_2$ terms can avoid instability when the means and variances are close to zero. An average MS-SSIM is then determined over all the test images.

\subsection{Single-user Transmission Performance}
The comparison between SIA-MIMO and the JPEG-Capacity, JPEG2000-Capacity, BPG-Capacity benchmarks and SIA-SC channel-equalization scheme is shown in Fig. \ref{SNR-MSSSIM MIMO}. Fig. \ref{SNR-MSSSIM MIMO}(a) shows the MS-SSIM results of the Open Images test set with CBR constraint $R = 1/24$ in $2\times2$ MIMO scenario. As can be seen, the SIA-SC MIMO-SVD scheme outperforms the separation-base benchmark in all SNRs. This is mainly benefited from the DeepJSCC architecture of the semantic communication system. The semantic communication system considers the main impact of channel noise on source semantics during the transmission process. The training process tries to maximally avoid the loss of source semantics caused by the channel. 

The advantages of the SIA-SC MIMO-SVD scheme become more significant as the number of antennas increases. In a $2\times2$ MIMO scenario, our scheme outperforms the SIA-SC scheme by over 2 dB. In a $4\times4$ MIMO scenario, our scheme exceeds the SIA-SC scheme by over 3 dB. In a large-scale 64x64 MIMO scenario, our scheme surpasses the SIA-SC MIMO-channel equalization scheme by over 10 dB.  Additionally, the proposed method outperforms traditional separate transmission schemes in both small-scale and large-scale antenna scenarios. However, simply applying an end-to-end semantic communication system to a multi-antenna system (SIA-SC MIMO-channel equalization) would not provide additional gains, and may even underperform separate transmission schemes as shown in Fig. \ref{SNR-MSSSIM MIMO}(b)(c)(d)(e)(f). 

A visual comparison of the reconstructed images for the SIA-SC MIMO-SVD scheme and the traditional scheme is presented in Fig. \ref{vis}. From top to bottom are scenes with different antenna numbers and SNR, and from left to right are the proposed SIA-SC MIMO-SVD scheme, BPG-Capacity scheme, and JPEG2000-Capacity scheme. It can be seen that BPG-Capacity and JPEG2000 produce visible blocking artifacts, especially in channels with antenna numbers and low SNR, which are not present in the images transmitted with SIA-SC. The visualized examples of the proposed scheme show that the impact of the channel on the image is a global faintening effect, still retaining details, while the comparative schemes show that the impact of the channel on the image is more of a blurring of details.

\subsection{Muti-user Transmission Performance}
The comparison between SIA-MIMO and the JPGE2000-Capacity and BPG-Capacity benchmarks is shown in Fig. \ref{MU SNR-MSSSIM MIMO}. Specifically, JPGE2000-Capacity and BPG-Capacity schemes calculate the channel capacity of two users using the NOMA scheme, which is the maximum number of bits that can be transmitted. Then the source information is compressed to the maximum bit number using BPG and JPEG source coding schemes, respectively. 

Since NOMA-Capacity \cite{cover1999elements, islam2016power} assumes the calculation of channel capacity under perfect successive interference cancellation (SIC), the interference between users becomes less significant compared to channel noise interference at lower SNR. Therefore, the performance of the two users under NOMA is comparable.
As can be seen from Fig. \ref{MU SNR-MSSSIM MIMO}, the performance of users allocated with higher power in the multi-user multi-antenna system is mostly similar to the BPG-Capacity and JPEG2000-Capacity schemes at high SNR. At low SNR, the proposed scheme has a distinct advantage. For example, in a $2\times2$ MIMO scenario when the SNR is lower than -2dB, and in a $4\times4$ MIMO scenario when the SNR is lower than -10dB.  Additionally, it is worth noting that the performance of User 1 and User 2 under the SIA-SC scheme is similar in high SNR. However, as the SNR decreases, the performance of User 2 gradually becomes inferior to that of User 1. This is because the performance of the semantic SIC scheme depends on the performance of User 1's recovery. When User 1's performance is poor, the error $\triangle_1$ in Eq. (\ref{error}) will be larger, reducing the recovery performance for User 2.

\subsection{Validation of Semantic System Analysis}
\subsubsection{SU-SISO scenario}
As defined in Eq. (\ref{SID_su_siso}) of the semantic performance analysis in Section \uppercase\expandafter{\romannumeral3}, it first calculates the average SID and MS-SSIM values for all test sets and uses a linear function to fit them, as shown in Fig. \ref{SID_fit},
\begin{equation}
MS-SSIM=a_1\times SID+a_2,
\label{MSSSIM_SID}
\end{equation}
where $a_1=-0.044, a_2=0.956$ in our trained system under the experimental setting.  By utilizing the SID expressions derived for the three scenarios (SU-MIMO, MU-SISO, and MU-MIMO) in Section \uppercase\expandafter{\romannumeral3}, the transmitter can directly use the expression fitted above between the SU-SISO semantic performance and SID to estimate the corresponding semantic performance of each user.

As shown in Fig. \ref{SOP_SISO_SU}, the dashed line represents the SOP for the SU-SISO scenario obtained through Eq. (\ref{SOP_su_siso}) and Eq. (\ref{Gaussian_sop}). At the same time, the discrete points are derived from simulations, which indicates that Eq. (\ref{SOP_su_siso}) and Eq. (\ref{Gaussian_sop}) provide an approximate closed-form expression for the SOP in the SU-SISO scenario.

\subsubsection{SU-MIMO scenario}
Fig. \ref{SU MIMO predict simulation} compares the predicted and simulated results for a single-user MIMO scenario, where the predicted line is obtained by calculating the SID for a single-user through Eq.  (\ref{SID_su_mimo}) and then calculating the corresponding semantic performance (MS-SSIM) by substituting it into Eq. (\ref{MSSSIM_SID}). As can be seen from Fig. \ref{SU MIMO predict simulation}, the trend of the prediction and simulation remains generally consistent. The predicted values are slightly higher than the simulated values, and the error gradually increases with more antennas. Overall, the transmitter can predict the performance of the image at the receiver given the known SNR. 

As shown in Fig. \ref{SOP_MIMO_SU}, the dashed line represents the SOP for the SU-MIMO scenario obtained through Eq. (\ref{SOP_su_mimo}) and Eq. (\ref{Gaussian_sop_su_mimo}). In this simulation, an interruption occurs when the semantic performance of MS-SSIM is less than 0.9. As the number of antennas increases, the probability of outage becomes smaller.

\subsubsection{MU-SISO scenario}
When considering multi-user scenarios, we first validate the advantage of the semantic SIC method (decode$\rightarrow$ encode$\rightarrow$ cancel). As shown in Fig. \ref{without_with_semantic_SIC}(a), the horizontal axis represents User 1's performance, and the vertical axis represents User 2's SID. It can be seen that using the semantic SIC scheme can significantly reduce User 2's SID under poor channel conditions. Specifically, as shown in Fig. \ref{without_with_semantic_SIC}(b), at low SNR, the quality of the received signal is poor. However, through the denoising ability of the semantic model itself, the semantic SIC method can eliminate certain noise. Therefore, using the semantic SIC scheme improves User 2's performance. 

As described by Eq. (\ref{SID_TE}) in Section \uppercase\expandafter{\romannumeral3}, the SIC after semantic SIC is related to the original SIC of User 1. As shown in Fig. \ref{without_with_semantic_SIC}(c), the horizontal axis represents the SIC of User 1, i.e. $SID_1$ in Eq. (\ref{SID_TE}), and the vertical axis represents the SID after semantic SIC, i.e. $SID_{sic}=\sum_{i=1}^K\mathbf{w}[i]|\mathbf{s}_1[i]-\mathbf{s}_{1,sic}[i]|^2$ in Eq. (\ref{SID_TE}). As can be seen from Fig. \ref{without_with_semantic_SIC}(c), when $SID_1$ is 0, SID SIC is slightly greater than 0. This is because the semantic model is a lossy encoding process, so there is still a loss even when the channel is noiseless. However, as the value of $SID_1$ increases, $SID_{sic}$ also slowly increases. Therefore, a linear function is used to fit the function $te$,
\begin{equation}
SID_{sic}=te(SID_1)=b_1\times SID_1+b_2,
\label{te_fit}
\end{equation}
where $b_1=0.64, b_2=0.21$ in our trained system under the experimental setting. Since there is no closed-form solution in theory to mathematically describe the performance of AI models, fitting is a necessary step. However, the above formula mainly shows the proportional relationship between SID and MS-SSIM, where the fitted parameters $b_1, b_2$ apply to the model under this performance. As long as the performance remains the same, even if the model or dataset changes, these parameters are still applicable. 

Fig. \ref{mu_siso} compares the predicted and simulated results for the two-user SISO scenario, where the predicted line is obtained by calculating the SID through Eq.  (\ref{SID_mu_siso_1}) and Eq. (\ref{SID_mu_siso_2}) and then calculating the corresponding semantic performance (MS-SSIM) by substituting it into Eq. (\ref{MSSSIM_SID}). As shown in Fig. \ref{mu_siso}, the predicted semantic performance values are slightly higher than the actual simulated values, but the overall trend is consistent and the gap is not significant. 

As shown in Fig. \ref{SOP_SISO_MU}, the theoretical values and simulation values of the SOP of user 1 and user 2 under the expected performance (MS-SSIM) of 0.9 and 0.7 respectively are plotted. The simulation values are basically on the theoretical curve, verifying the feasibility of the closed-form expression of the SOP of Eq. (\ref{SOP_mu_siso_1}), Eq. (\ref{Gaussian_sop_mu_siso}), Eq. (\ref{SOP_mu_siso_2}) and Eq. (\ref{Gaussian_sop_mu_siso_2}) in the MU-SISO scenario.

\subsubsection{MU-MIMO scenario}
Fig. \ref{MU_MIMO_Simulation_predict} shows the predicted and simulated curves for the MU-MIMO scenario. As with other validation methods, it first calculates the SID for the two users respectively using Eq. (\ref{mu_mimo_sid_1}) and Eq. (\ref{mu_mimo_sid_2}), and then calculates the corresponding performance using Eq. (\ref{MSSSIM_SID}). As with other scenarios, the predicted values are slightly greater than the simulated values, with the overall trend remaining consistent and the errors being relatively small.

Fig. \ref{SOP_MIMO_MU}(a)(b)(c) represent SOP for the 4x4, 8x8, and 16x16 MU-MIMO scenarios, with an MS-SSIM target value set at 0.9. The simulation values closely align with the theoretical curve, with a slightly larger error observed in the 16x16 scenario, although it remains within an acceptable range.

\section{Conclusion}
This paper presents a semantic importance-aware based communication system (SIA-SC) over MIMO Rayleigh fading channels. Combining the semantic symbols' inequality and the equivalent subchannels of MIMO channels based on SVD maximizes the end-to-end semantic performance through the new layer mapping method. For multi-user scenarios, a method of semantic interference cancellation is proposed. Moreover, a novel metric called SID has been introduced to provide a unified representation of semantic performance, which is affected by CBR and SNR.  With the help of the proposed metric, we derived performance expressions and SOP of SIA-SC for SU-SISO, SU-MIMO, MU-MIMO and MU-MIMO scenarios. Numerical experiments show that SIA-SC can significantly improve semantic performance across a large variety of scenarios.


\section*{Acknowledgment}
This work is supported in part by the National Key Research and Development Program of China under Grant 2022YFB2902102.

\small
\bibliography{IEEEabrv, reference}

\end{document}